\documentclass[showkeys,reprint,superscriptaddress,nofootinbib,amsmath,amssymb,aps,floatfix,]{revtex4-1}
\usepackage[latin1]{inputenc}
\usepackage{float}
\usepackage{amsmath}
\usepackage[english]{babel}
\usepackage{amsfonts}
\usepackage{amssymb}
\usepackage{fancyhdr}
\usepackage{esdiff}
\usepackage{comment}
\usepackage{textcomp}
\usepackage{cancel}
\usepackage{tikz} 
\usepackage{subfigure}
\usepackage{graphicx}
\usepackage{stackengine}
\usepackage[font=small]{caption}
\usepackage{epsfig}
\usepackage{epstopdf}
\usepackage[T1]{fontenc}
\usepackage{lmodern}
\usepackage{ctable}
\usepackage{color, colortbl}
\usepackage[font=small,labelfont=bf,justification=raggedright]{caption}
\definecolor{Gray}{gray}{0.9}
\definecolor{LightCyan}{rgb}{0.88,1,1}

\newcommand*{\avk}{\langle k \rangle}

\begin{document}
\title{Reactive random walkers on complex networks}

\author{Giulia Cencetti}
\affiliation{Dipartimento di Ingegneria dell'Informazione, Universit\`{a} degli Studi di Firenze}
\affiliation{Dipartimento di Fisica e Astronomia, Universit\`{a} degli Studi di Firenze, INFN and CSDC}

\author{Federico Battiston}
\affiliation{Department of Network and Data Science, Central European University, Budapest 1051, Hungary}

\author{Duccio Fanelli}
\affiliation{Dipartimento di Fisica e Astronomia, Universit\`{a} degli Studi di Firenze, INFN and CSDC}

\author{Vito Latora}
\affiliation{School of Mathematical Sciences, Queen Mary University of
	London, London E1 4NS, UK}
\affiliation{Dipartimento di Fisica ed Astronomia, Universit\`a di Catania and INFN, I-95123 Catania, Italy}

\begin{abstract}

We introduce and study a metapopulation model of random walkers
interacting at the nodes of a complex network. The model integrates
random relocation moves over the links of the network with local
interactions depending on the node occupation probabilities. The model
is highly versatile, as the motion of the walkers can be fed on
topological properties of the nodes, such as their degree, while any
general nonlinear function of the occupation probability of a node can
be considered as local reaction term. In addition to this, the
relative strength of reaction and relocation can be tuned at will,
depending on the specific application being examined. We derive an
analytical expression for the occupation probability of the walkers at
equilibrium in the most general case. We show that it depends 
on different order derivatives of the local reaction functions and
not only on the degree of a node, but also on the average degree of
its neighbours at various distances. For such a reason, reactive
random walkers are very sensitive to the structure of a network and
are a powerful way to detect network properties such as symmetries or
degree-degree correlations.  As possible applications, we first
discuss how the occupation probability of reactive random walkers can
be used to define novel measures of functional centrality for the
nodes of a network. We then illustrate how network components with the
same symmetries can be revealed by tracking the evolution of reactive
walkers.  Finally, we show that the dynamics of our model is
influenced by the presence of degree-degree correlations, so that
assortative and disassortative networks can be classified by
quantitative indicators based on reactive walkers.
\end{abstract}	

\maketitle

\section{Introduction}

%
The architecture of various social, biological and man-made systems
composed by many interacting elements can be well described in terms
of complex networks \cite{Newman10,LatoraNicosiaRusso17book}.  The
ability of all such systems to execute complex tasks and implement
dedicated functions is indeed intimately connected to their underlying
architecture. Studies on epidemic spreading, synchronization and game
theory have shown how different network topologies can affect the
emergence and the properties of collective behaviours in a given system
\cite{BarratBarthelemyVespignani08}.  Similarly, ingenious techniques
have been proposed to reconstructing the topology of a given network
from direct inspection of its emerging dynamics\cite{ChengShen10,DeDomenico17,AsllaniCarlettiDiPattiFanelliPiazza18}. Fully understanding the
interplay between structure and function is generally considered today
as one of the grand challenges of network science.

%
Random walks are probably the simplest among the many dynamical
processes which have been studied on networks. Since the pioneering
works of Pearson~\cite{Pearson05}, who also coined the term, random
walks have been extensively investigated in different fields ranging
from probability theory to statistical physics and computer science,
and have found a number of practical applications. A random walk on a
network iinvolves an agent that performs local hops from one node to one of
its neighbours, producing in this way random sequences of adjacent
nodes~\cite{AldousFill02,NohRieger04,Yang05}. Despite the simplicity
of the process, random walks have been proven a fundamental tool to
unravel unknown features of the underlying
network~\cite{Newman10,GomezGardenesLatora08,ChengShen10,LinZhang14,DeDomenico17}.
For instance, they have been used to identify the most central
nodes~\cite{LatoraNicosiaRusso17book,Newman10,BrinPage98,Bonacich72,Klemm_etal12,Iannelli_etal18}
or the modules of a given network
\cite{Rosvall_2008,Djurdjevac_12,Sarich_14}. The trajectories of
random walkers have also turned useful to uncover hidden relationships
between nodes of the network, like symmetries or degree-degree 
correlations~\cite{PeelDelvenneLambiotte17}.  More in general, random
walks on complex networks are considered to be at the heart of several
real-world dynamical systems, like diseases
spreading~\cite{Iannelli17}, financial markets~\cite{Azoff94},
decision-making in the brain~\cite{Wang02,Kamienkowski_etal11},
foraging of animals~\cite{Schippers_etal96}, innovation growth~\cite{Iacopini_18,Conrad_etal18} and more. They have also
found applications in the context of metapopulation
models~\cite{Levins69,Hanski98,Ovaskainen01,UrbanKeitt01,NicosiaBagnoliLatora11,CencettiBagnoliDiPattiFanelli15},
where the nodes of the network represent discrete patches occupied by
members of a local population, and the random walk process describes
the migration from patch to patch.

%
In the simplest possible case, at each time step, a random walker jumps
from one node to one of its first neighbours, 
which is chosen at random with uniform
probability. However, the process can 
be generalized so as to bias the walk towards nodes that display
specific features. In the case of degree-biased random walkers,
for instance, the transition probability between two adjacent nodes
is gauged by the degree $k$ of the target node. This can be done 
so as to impose a preferential movement
towards hubs or, alternatively, towards poorly connected
nodes~\cite{GomezGardenesLatora08}. The versatility of this type of random
walks has inspired in the last years an abundance of methods to investigate
the network structure of real-world systems~\cite{BonaventuraNicosiaLatora14}. Biased random walks have also
been employed for community
detection~\cite{ZlaticGabrielliCaldarelli10}, to define new centrality
measures~\cite{Lee09,Delvenne11}, to characterize the structure of
multi-layer networks~\cite{BattistonNicosiaLatora16} and to measure
degree-degree correlations~\cite{GomezGardenesLatora08,BaronchelliPastorSatorras10,
  Burda09, SinatraGomexGardenesLambiotteNicosiaLatora11}.

Random walks are usually depicted as models for diffusion. It is
however important to distinguish among such two related but different
concepts. More specifically, diffusion refers to the flow of a
(material or immaterial) substance, on a continuous or discrete
support, from regions of high concentration to regions of low
concentration. This process inevitably yields a space-homogeneous
redistribution of the density, which is forcefully subject to detailed
balance constraints. When diffusion occurs on a network, the system
evolves towards an asymptotic state where all nodes are equally
populated, often termed as consensus~\cite{DeGroot74}. Hence, the
stationary state associated to a purely diffusive process does not
bear information on the underlying network structure, which is solely
influencing the dynamics during the transient, before consensus is
eventually reached. The stationary distribution as attained by random
walkers on a network is instead proportional to the connectivity of
the nodes, and this basic fact hints at how they can prove more
informative than diffusion when the focus is on the
network topology~\cite{AsllaniCarlettiDiPattiFanelliPiazza18}.

%
While random walks are the basic ingredient to describe mobility, they
do not take into account of the possibile interactions between agents present
in the same node of a
network. These are typically described by a local dynamics, which can
be different for each node. Local dynamics have been frequently coupled
with diffusive processes to describe the self-consistent evolution of
mutually coupled species, when subject to the combined influence of
diffusion and reaction terms~\cite{Murray02,OthmerScriven71,
  OthmerScriven74, NakaoMikhailov10,
  AsllaniChallengerPavoneSacconiFanelli14,
  AngstmannDonnellyHenry13}. In this work, we propose a model of
\textit{reactive random walkers}, where generalized biased random walkers not
only navigate the system but also interact when they meet at the
nodes of the network. At variance with conventional diffusion, 
in reactive random walkers the 
probability of relocation between adjacent nodes is also sensitive to
local reactions, which ultimately confer to each node a
self-identity. For such a reason, the occupation probability of a
given node depends not only on the connectivity pattern but also on
the ability of the node itself to attract walkers. This last property
can be tuned at will by properly shaping the reaction term, and this
enables in turn to highlight different characteristics of the network
structure. Reactive random walkers are highly versatile and motivate a
series of applications aimed at uncovering the topology of the
discrete support where the dynamics takes place. In particular, in this
paper we will focus on: {\em (i)}
the definition of a novel functional centrality measure, {\em (ii)}
the issue of revealing hidden symmetries in a graph 
and {\em (iii)} the problem of characterizing node degree-degree
correlations in complex networks.

%
The article is organised as follows. Section \ref{sec_model}
introduces our model of reactive random walkers in its most general
form.  Examples on a number of small graphs are reported to elucidate
the main ingredients of the model both at the level of the choice of
the reaction functions and of the type of bias in the walk.  In
Section \ref{sec_analytic} we analytically derive the stationary state
of the dynamics of the model by means of the perturbative calculation.
In Section
\ref{sec_centr} we elaborate on a novel measures of functional
centrality, as a first application of the model. Section
\ref{sec_symm} unveils the relationship between reactive random
walkers and network symmetries. In Section \ref{sec_degreecorr} we
further investigate the connection between dynamics and structure, by
proposing an alternative indicator of degree-degree correlations in
networks.  Finally, in Section \ref{sec_concl} we discuss
possible further extensions of the proposed model.

\section{Model}
\label{sec_model}

Our model describes the dynamics of {\em reactive random walkers},
i.e.  random walkers moving over the links of a complex network and
interacting at its nodes.  Let us consider an undirected and
unweighted network with $N$ nodes and $K$ edges, described by a
symmetric adjacency matrix $A = \{a_{ij}\}$, where $a_{ij}=1$ if nodes
$i$ and $j$ are linked, and $a_{ij}=0$ otherwise. We denote as $x_i
(t)$ the occupation density, at time $t$, of node $i$, with
$i=1,2,\ldots, N$, so that the state of the entire network at time $t$
is completely described by the vector $\boldsymbol{x}(t)= (x_1(t),
x_2(t), \ldots, x_N(t))$. The occupation density $\boldsymbol x$ 
shall be normalised as $\sum_i x_i(t)=1\ \forall t$, so that it can be considered
as an occupation probability. 
The law governing the time evolution of
$\boldsymbol{x}(t)$ takes into account the network topology, i.e. the
adjacency matrix $A$, and also the specific characteristics of each
individual node through a set of local reaction functions.  This is
formally expressed by the following $N$ equations:
\begin{equation}
\dot x_i = (1-\mu) f(x_i) + \mu\sum_{j=1}^N l^{\rm RW}_{ij}x_j  \hspace{5mm} i=1,...,N
\label{gamma_dynRW}
\end{equation} 

where $\mu$ is a tuning parameter, thereon referred to as the
\textit{mobility parameter}, which takes values in $[0,1]$ and
enables us to modulate the weight of two contributions.  The first
term on the right-hand side of Eqs.~\eqref{gamma_dynRW} accounts for
the local reaction at each node $i$, and is ruled by a function
$f(x_i)$ of the occupation probability $x_i$.  For simplicity we
assume that the reaction function $f$ is the same for all nodes.
The second term takes into account the topology of the network and
describes the mobility on it by means of the {\em random walk
	Laplacian} $L^{\rm RW}= \{ l^{\rm RW}_{ij} \}$.  This Laplacian is
defined as:
\begin{equation} 
l^{\rm RW}_{ij}=\pi_{ij}-\delta_{ij}
\label{laplacian_rw}
\end{equation}
where $\Pi = \{ \pi_{ij} \}$ is the transition matrix of a random
walk. Entry $\pi_{ji}$ of matrix $\Pi$ represents the probability 
of the random walker to move from node $i$ to node $j$ (see Appendix
A). Notice that $\sum_j \pi_{ji}=1$ $\forall i$.  In the simplest
possible case we can assume that the random walk is unbiased. This
means that the probability of leaving node $i$ is equally distributed
among all its adjacent nodes $j$, so that we can set $\pi_{ji}= a_{ij}
/ k_i$ for each $j$. Here, we consider instead a more general
transition matrix in the form:
\begin{equation} 
\pi_{ji} =  \frac{ a_{ij} k_j^{\alpha} } {\sum_l a_{il}  k_l^{\alpha} }
\end{equation}
which describes degree-biased random walks, i.e. random walkers 
whose motion also depends on the degree of the node $j$, and such a
dependence can be tuned by changing the value of the exponent
$\alpha$~\cite{GomezGardenesLatora08}.
Namely, for $\alpha>0$, the walker at node $i$ will preferentially
move to neighbours with high degree while, for $\alpha<0$, it will
instead prefer low degree neighbours. Finally, for $\alpha=0$,
we recover the transition matrix $\pi_{ji}= a_{ij}/k_i$
of the standard unbiased random walk. 

Summing up, the main ingredients and tuning parameters of the reactive
random walkers model in Eqs.~\eqref{gamma_dynRW} 
are: the network topology, encoded in the
adjacency matrix $A$ of the underlying mobility graph; the bias
parameter $\alpha \in \mathbb{R}$, which allows to explore the graph in
different ways; the local reaction functions ruling the interactions
at nodes; and the mobility parameter $\mu \in [0,1]$ to weight the relative
strength of reaction and relocation. 
Notice that the model of reactive random walkers we have introduced 
recalls metapopulation
models~~\cite{Levins69,Ovaskainen01,NicosiaBagnoliLatora11}, for which the occupation probability of each node of the network wherein the population is allocated is governed by a random walk process,
as well as by a local term accounting for birth and death on each
environment. Eqs.~\eqref{gamma_dynRW} are also similar to those
describing reaction-diffusion processes, but where $x_i(t)$ represents
the density at node $i$ at time $t$, and the Laplacian 
matrix $L^{\rm RW}$ of Eq.~\eqref{laplacian_rw}
is replaced by the matrix
$L^{\rm Diff} = \{ l_{ij}^{\rm Diff} \}$ that stems from a purely diffusive process.
For similarities and differences between the two definitions of Laplacian see
Appendix A.

\begin{figure}
	\captionsetup{justification=raggedright,singlelinecheck=false}
	\subfigure[]{\includegraphics[width=5cm]{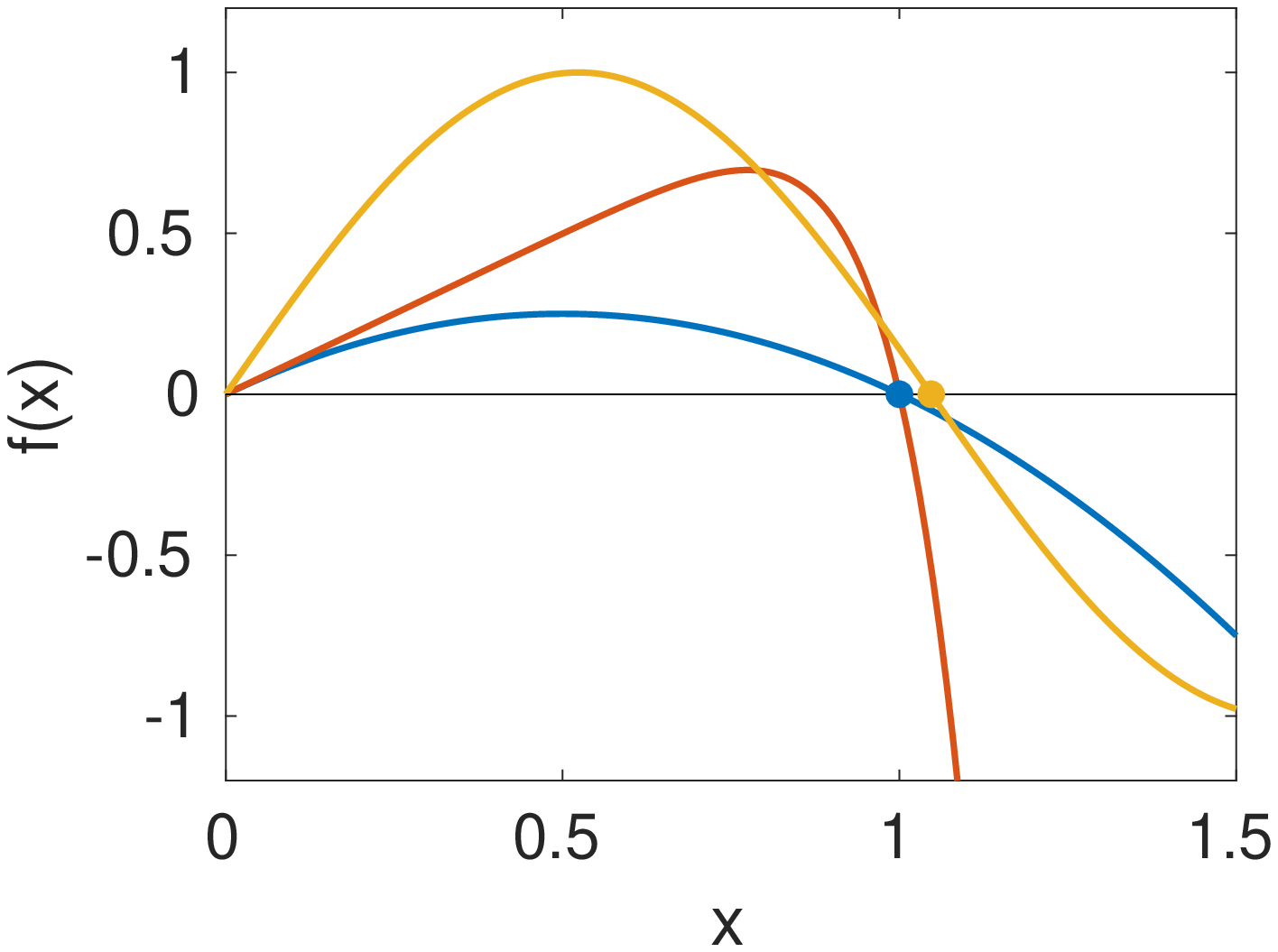}
		\label{fig:react}}
	\subfigure[]{\begin{tikzpicture} [xscale=.9,
		yscale=.8] \draw[fill] (1.6,0) circle
		[radius=0.05]; \node [right] at (1.6,0) {1};
		\draw[fill] (1.4,-1.2) circle [radius=0.05]; \node
		[right] at (1.4,-1.2) {5}; \draw[fill]
		(0.2,-1.2) circle [radius=0.05]; \node
		[below] at (0.2,-1.2) {4}; \draw[fill] (-0,0)
		circle [radius=0.05]; \node [left] at (-0,0)
		{3}; \draw[fill] (.8,.6) circle
		[radius=0.05]; \node [above] at (0.8,.6) {2};
		\draw[fill] (1.4,-1.9) circle [radius=0.05];
		\node [right] at (1.4,-1.9) {6}; \draw[fill]
		(1.4,-2.5) circle [radius=0.05]; \node
		[right] at (1.4,-2.5) {7}; \draw[fill]
		(.8,1.7) circle [radius=0.05]; \node
		[left] at (.8,1.7) {8}; \draw[fill]
		(.8,-.6) circle [radius=0.05]; \node
		[below] at (.8,-.6) {9}; \draw [-]
		(1.6,0) -- (.8,0.6); \draw [-] (1.6,0)
		-- (.8,-.6); \draw [-] (1.6,0) --
		(.8,1.7); \draw [-] (1.6,0) --
		(1.4,-1.2); \draw [-] (0,0) --
		(.8,0.6); \draw [-] (0,0) -- (.8,-.6);
		\draw [-] (0,0) -- (.8,1.7); \draw [-]
		(0,0) -- (0.2,-1.2); \draw [-]
		(1.4,-1.2) -- (1.4,-1.9); \draw [-]
		(1.4,-2.5) -- (1.4,-1.9); \draw [-]
		(0.2,-1.2) -- (1.4,-1.2);
		\end{tikzpicture}
		\label{fig_net9nodes}}
	\subfigure[]{\includegraphics[width=9cm]{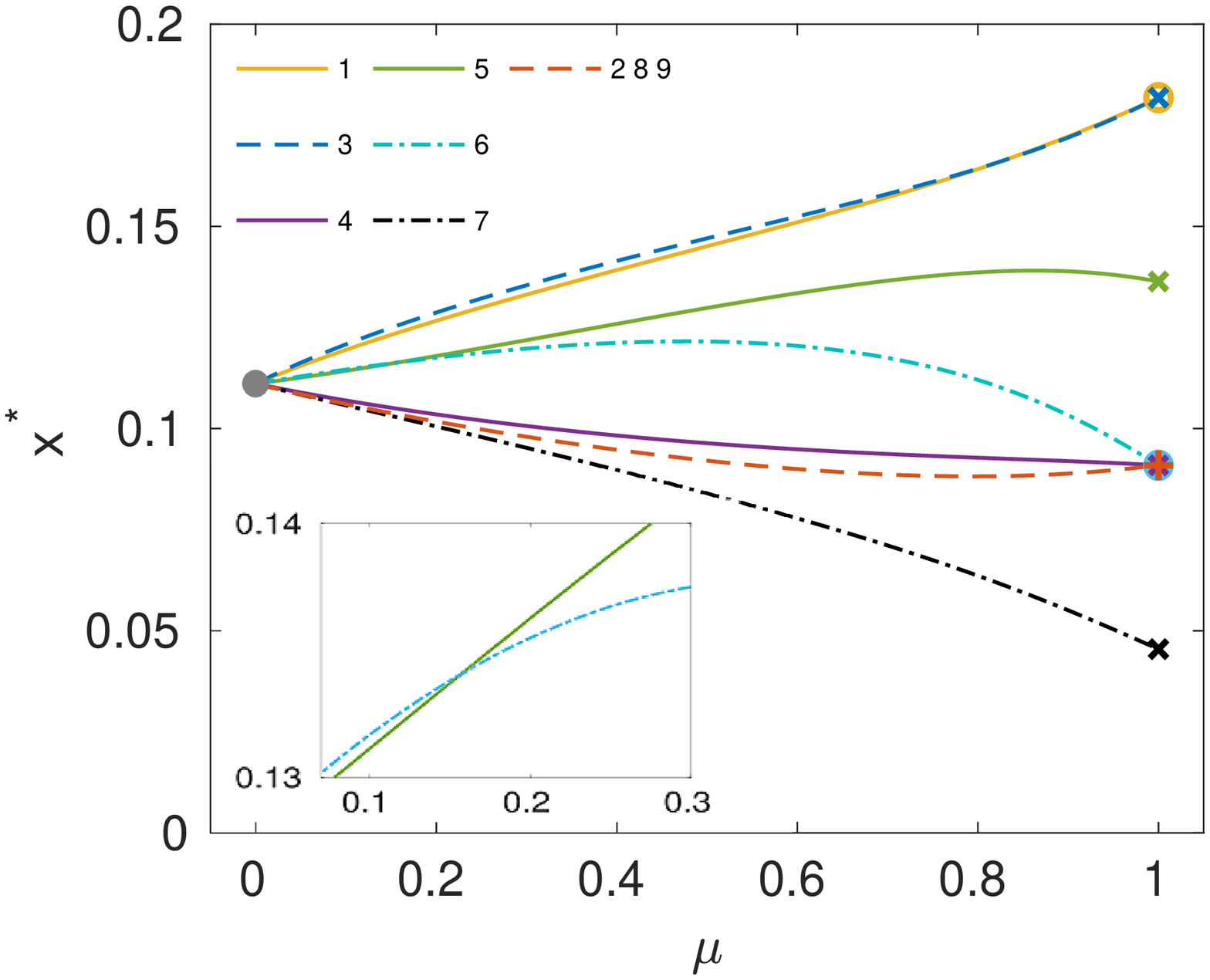}
		\label{fig_x9nodes}}
	\caption{ (a): Examples of possible reaction functions to be used in
		Eqs.~\eqref{gamma_dynRW}: $f(x)= x-x^2$ in blue, $f(x)= x-x^{10}$ in
		red and $f(x)= \sin(3x)$ in yellow. (b) A graph of $N=9$ nodes, and
		(c) the fixed point $\boldsymbol{x^*}$ obtained when a reactive
		random walk model with $f(x)=x-x^2$ and different values of the
		mobility parameter $\mu$ is implemented on such a graph. The inset
		represents a zoom showing the inversion of $x^*_5$ and $x^*_6$
		obtained by changing $\mu$.}
	\label{fig_9nodes}
\end{figure}

\medskip
{\bf ~  Limiting case $\mu=0$.~~~}	
Let us begin the analysis of the reactive random walk model by
considering its two limits, namely $\mu=0$ and $\mu=1$. In the
first limit, the mobility is completely suppressed and the dynamics of
each node is independent of the others. Since we have assumed that the
function $f$ is the same for each node, Eqs. \eqref{gamma_dynRW}
reduce to solve the 1-dimensional system $\dot x=f(x)$. 
In principle the reaction function $f$ can be freely chosen among all
the functions $f$: $\mathbb{R}\rightarrow\mathbb{R}$. However 
interesting cases are found when the variable $x(t)$ is bound to converge
towards a stationary point, $x^*$, defined by $f(x^*)=0$. The
function $f$ should then be chosen among the continuous functions and
such that $0$ is included in its image. Moreover, in order to have
equilibrium stability, it is necessary that $f$ is monotonically
decreasing in, at least, one of the points where it vanishes, in order to
ensure that there exists (at least) one stable fixed point $x^*$. Some possible 
examples of reaction functions are reported in Fig. \ref{fig:react}.

\medskip
{\bf ~ Limiting case $\mu=1$.~~~}	
In the opposite limit, when the mobility parameter takes its maximum
value $\mu=1$, Eqs.~\eqref{gamma_dynRW} describe a pure random
walk process.
\begin{figure}
	\hspace{0.6cm} \subfigure[]{\begin{tikzpicture} [xscale=1.3,
		yscale=1.3] \draw[fill] (0,1.6) circle [radius=0.06]; \node
		[above] at (0,1.6) {1}; \draw[fill] (0.6,0) circle [radius=0.06];
		\node [below right] at (0.6,0) {4}; \draw[fill] (-0.6,0) circle
		[radius=0.06]; \node [below left] at (-0.6,0) {3};
		\draw[fill] (0,.7) circle [radius=0.06]; \node [above left]
		at (0,.7) {2}; \draw [-] (0,.7) -- (0,1.7); \draw [-]
		(-0.6,0) -- (0.6,0); \draw [-] (0,.7) -- (0.6,0); \draw [-]
		(0,.7) -- (-0.6,0);
		
		\end{tikzpicture}
	}
	\hspace{0.9cm}
	\subfigure[]{\includegraphics[width=4.27cm]{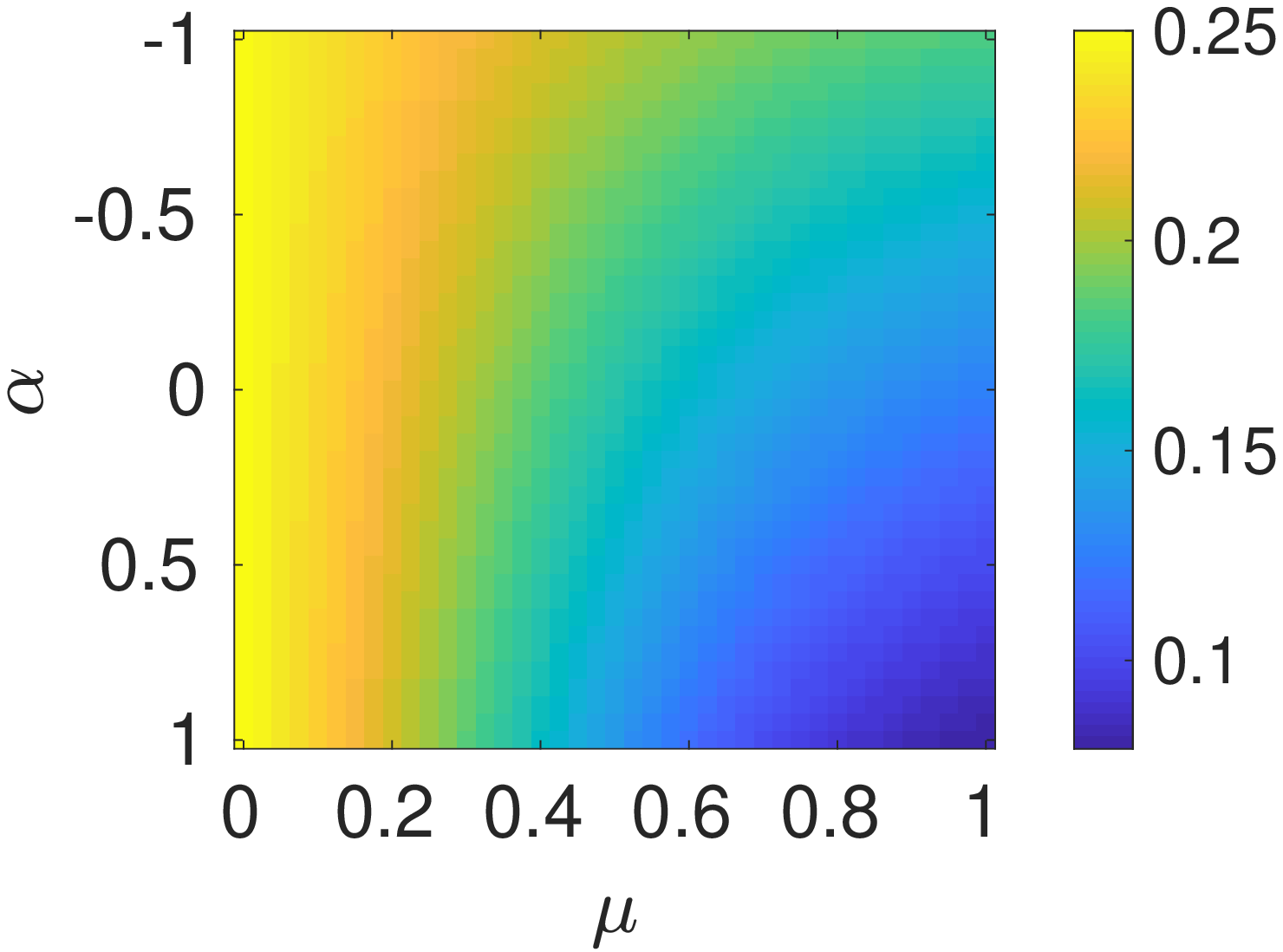}}
	\subfigure[]{\includegraphics[width=4.27cm]{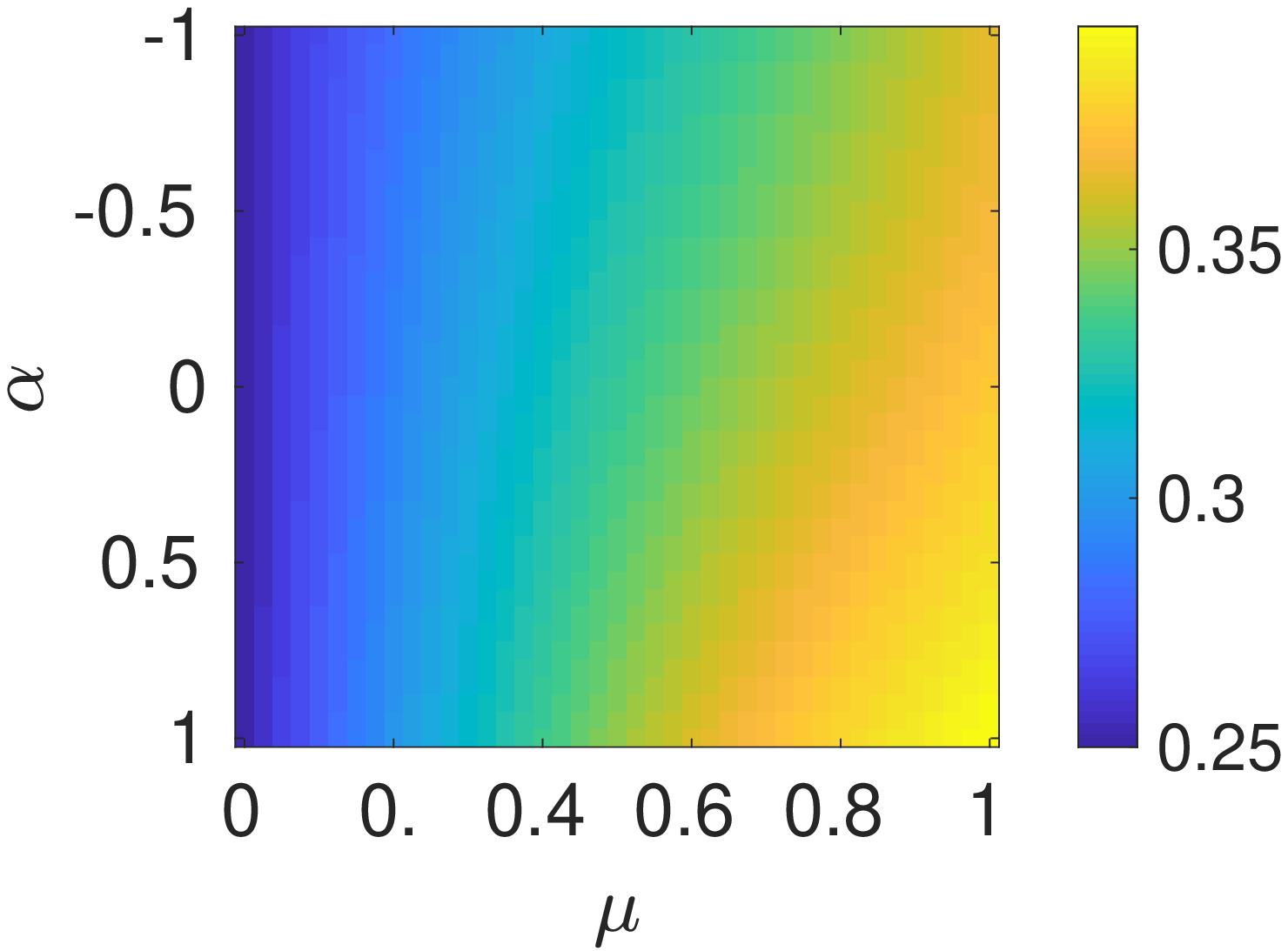}}
	\subfigure[]{\includegraphics[width=4.27cm]{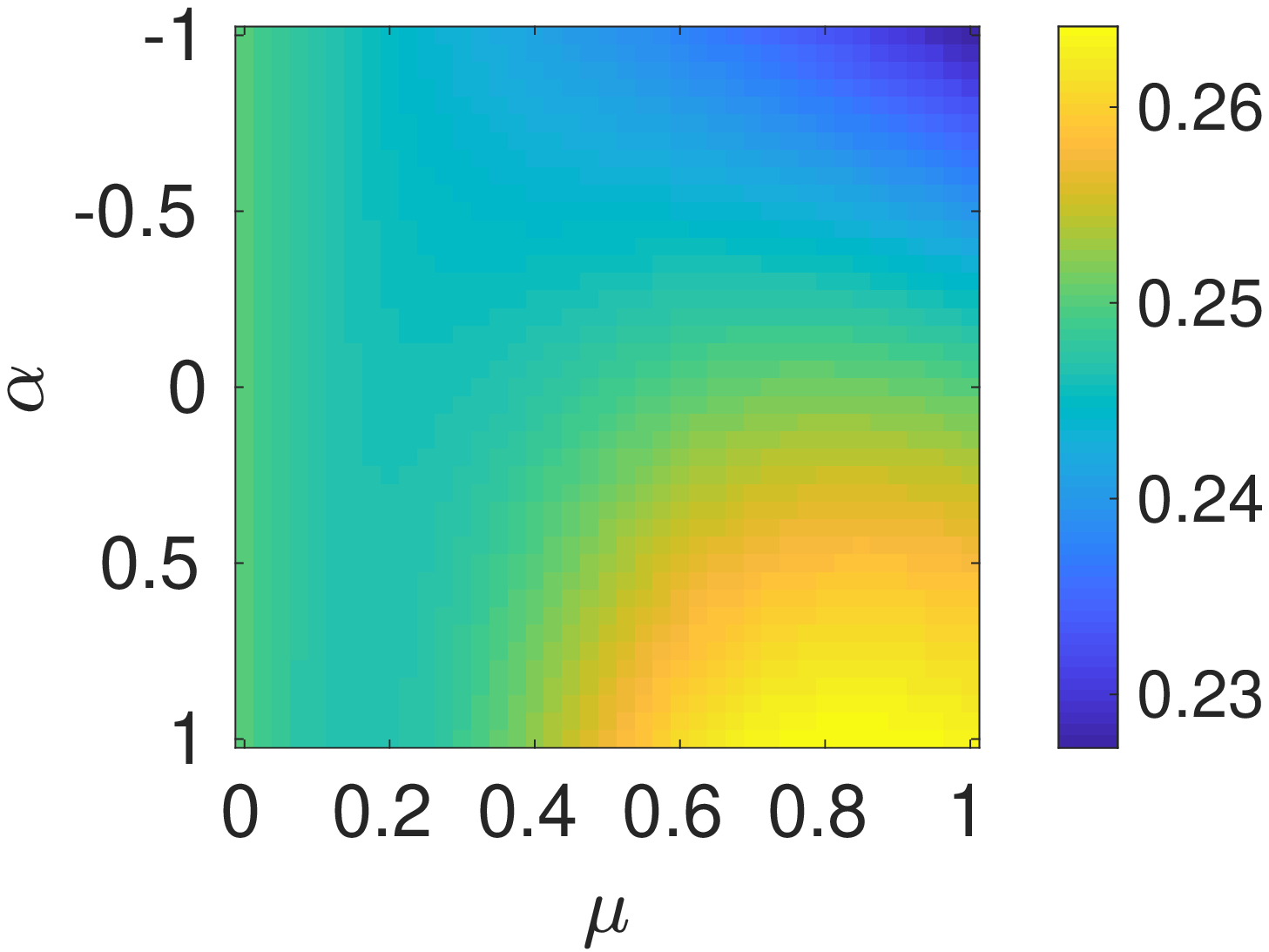}}
	
	\caption{Stationary occupation probability of the nodes of graph in
		(a) is shown for a reactive random walk as a function
		of the mobility parameter $\mu$ and of the bias exponent
		$\alpha$: node 1 (b), node 2 (c), nodes 3 and 4 (d).  The latter
		two nodes yield identical patterns (consequently displayed in just one
		figure), being symmetric nodes. The same reaction function as in
		Fig. \ref{fig_9nodes} has been chosen.}
	\label{fig:bias}
\end{figure}
The stationary distribution $\boldsymbol{x^*} = \{ x_1^*, x_2^*,\ldots, x_N^*\}$
of the dynamics
in this limit is obtained by $L^{\rm RW} \boldsymbol{x^*}=0$,
which is equivalent to
$\Pi \boldsymbol{x^*}=\boldsymbol{x^*}$. The
Perron-Frobenius~\cite{Perron07, Frobenius12} theorem ensures that, if
the graph is connected and contains at least one odd cycle, the fixed
point $\boldsymbol{x^*}$ always exists and is unique. In the case of 
degree-biased random walks we get~\cite{GomezGardenesLatora08}:
\begin{equation}
x_i^* = \frac{c_i k_i^{\alpha}}{\sum_l c_l k_l^{\alpha}} \text{ with } c_i = \sum_j a_{ij}k_j^{\alpha}.
\label{fp_bias}
\end{equation}
Such an expression, for $\alpha=0$, reduces to $x_i^*=k_i/2K$, meaning
that the walker, after a long enough period of time, is found on a
node $i$ with a probability linearly proportional to the node degree
$k_i$. In this case the asymptotic distribution is 
completely characterized by the degree $\boldsymbol{k}$ of the graph,
with better-connected nodes having a larger probability of being
visited by the walker.
\\
The general expression for the asymptotic distribution at a node $i$,
when $\alpha\neq0$, depends instead not only on the degree $k_i$ of node
$i$, but also on the degrees of the first neighbours of node $i$,
through the coefficient $c_i$, and such dependence can be tuned by
changing the value of the exponent $\alpha$. For instance, 
optimal values of the bias, which depend both on the degree distribution
and on the degree-degree correlations of a network, 
can be found to obtain maximal-entropy random walks
~\cite{GomezGardenesLatora08,Burda09,SinatraGomexGardenesLambiotteNicosiaLatora11} 
or to induce the emergence of synchronization~\cite{GomezGardenesNicosiaSinatraLatora13}.

\medskip	
{\bf ~ The general case.~~~}	

The most interesting dynamics of our model emerges at intermediate
values of the mobility parameter $\mu$, when interactions at nodes and
random movements between nodes are entangled. In this case, the
walkers move on the network jumping from node to node, so that the
node occupation probability depends on the network connectivity
because of the Laplacian contribution but, at the same time, it
evolves at each node according to the reaction function. Reaction
functions in turn depend on the occupation probability, so that we
have different contributions for differently populated node. This
leads to a stationary probability $\boldsymbol{x^*}$ reflecting the
topology of the graph in a way that is non trivial and worth
analysing.  The stationary probability of the model can be obtained,
for any value of $\mu$ in $[0,1]$, by setting $\dot x_i=0$ in
Eqs.~\eqref{gamma_dynRW} and solving numerically the following
recursive equations:
\begin{equation}
x_i^* = \sum_j \frac{a_{ij}}{k_j}x_j^* + \frac{(1 - \mu)}{\mu}
f(x_i^*).
\label{conv_eq}
\end{equation}	
Notice however that, when $\mu\neq 1$, the state $x_i(t)$ of node $i$ 
in Eqs.~\eqref{gamma_dynRW} is not constrained between 0 and 1. This
is an effect caused by the reaction term, which behaves as a source
term at each node. If we want to interpret the state of the network as
an occupation probability, we need then to further impose the
normalization, for instance we can consider the vector
$\boldsymbol{x}/\sum_ix_i$ instead of the vector $\boldsymbol{x}$.

In the following, we will consider a series of examples so as to get a
first insight on the properties of the stationary distribution
$\boldsymbol x^*$ for different network structures and for different
values of the two main tuning parameters of
the model, namely the mobility parameter $\mu$ and the bias exponent
$\alpha$.

%
%
In Fig. \ref{fig_9nodes}, as local interaction,    
we consider the logistic function 
$f(x)=x-x^2$ shown in panel (a), and we
implement the model on the graph of $N=9$ nodes displayed in panel
(b).  Panel (c) reports the obtained values of the components of the
normalized fixed point $\frac{1}{\sum_ix_i^*}(x_1^*, x_2^*,...,x_9^*)$
as functions of the mobility parameter $\mu$, when $\alpha$ is fixed
to zero.  The numerical results are in
agreement with the expected behaviours in the two limiting cases
$\mu=0$ and $\mu=1$.  In particular, we get $\boldsymbol
x^*=\boldsymbol k/2K$ for $\mu=1$, and $\boldsymbol x^*=\boldsymbol
1/N$ for $\mu=0$, where $\boldsymbol 1$ denotes an $N$-dimensional
vector with all entries being identically equal to 1. This means that
all the curves in the figure start from the same point $\boldsymbol x^*$ at $\mu
=0$, while for $\mu=1$ we observe four different points $\boldsymbol x^*$. The
graph considered has in fact nodes with four different degrees, namely
$k=1, 2, 3$ and 4, and curves corresponding to nodes with same number
of links will converge to the same point $x^*$ for $\mu=1$.
However, at intermediate values of $\mu$, even nodes with the same 
degree can exhibit different values of $x^*$ (with the exception of some
of them, see Section \ref{sec_symm} for a discussion on symmetric
nodes) going from their degree class at $\mu=1$ towards $1/N$ at
$\mu=0$.  In particular, the various curves of  $x^*$ as a function of
$\mu$ can cluster in a different way
when heading towards the limit $\mu = 0$. Let us 
focus for instance on the behaviour of the node 6 of the graph.
Such a node belongs to
the degree-2 class but, following the curve of its stationary state when
it goes from $\mu=1$ to $\mu=0$, we notice that it separates from the curves
of the other nodes of its class, approaching the curve of node 5, $x_5^*$,
although the latter node is characterized by a larger degree 
($k_5=3$). Moreover, node 6 even overcomes node 5 for small values of
$\mu$ before both curve collapse towards the homogeneous solution. The
crossing between the two curves is highlighted in the inset of
Fig. \ref{fig_x9nodes}.

%
In the most general case, in our model it is possible to tune both the
local dynamics, by choosing different reaction functions $f(x)$, and
the bias in the random walk, by considering values of the exponent $\alpha
\neq 0$. An illustrative example is reported in 
Fig. \ref{fig:bias} in the case of a smaller 
graph with only four nodes. The three coloured panels show the three different
values of the fixed point at the nodes of the 
network as functions of the mobility parameter $\mu$ and the bias exponent
$\alpha$. Notice that node 3 and 4 have the same symmetry in the graph, 
so they reach the same fixed point 
(see Section \ref{sec_symm} for a discussion of symmetries). 
In detail, while for $\mu$ and $\alpha$ equal to zero the four nodes
exhibit the same value of the occupation probability, $x^*_i=0.25$
$\forall i$, when we increase the mobility parameter we observe 
a non-trivial behaviour of these values, which in general decrease
for low-degree nodes and increase for high-degree nodes. 
The effect of introducing a degree-bias in the random walk 
by turning on and tuning the bias parameter is instead that the occupation probability of the
most connected nodes (see nodes 2, 3 and 4) is enhanced for positive
and decreased for negative values of $\alpha$. The opposite happens
for the less connected nodes (node 1).

\begin{figure}[h]
	\subfigure{\includegraphics[width=4.4cm]{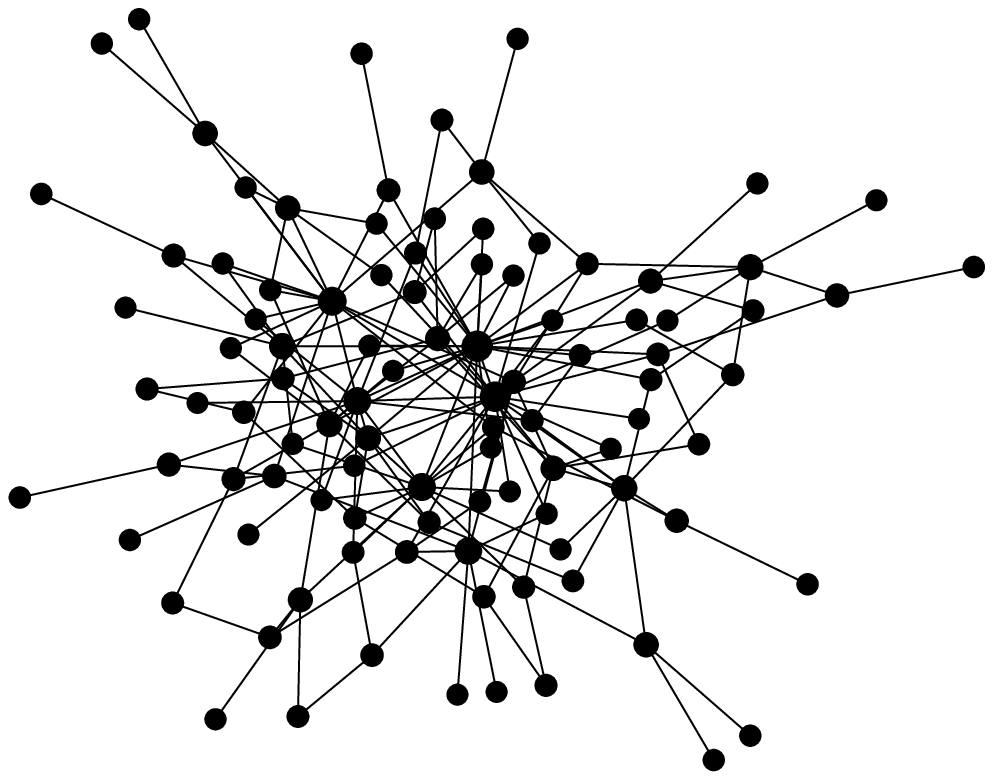}}
	\subfigure{\includegraphics[width=4.4cm]{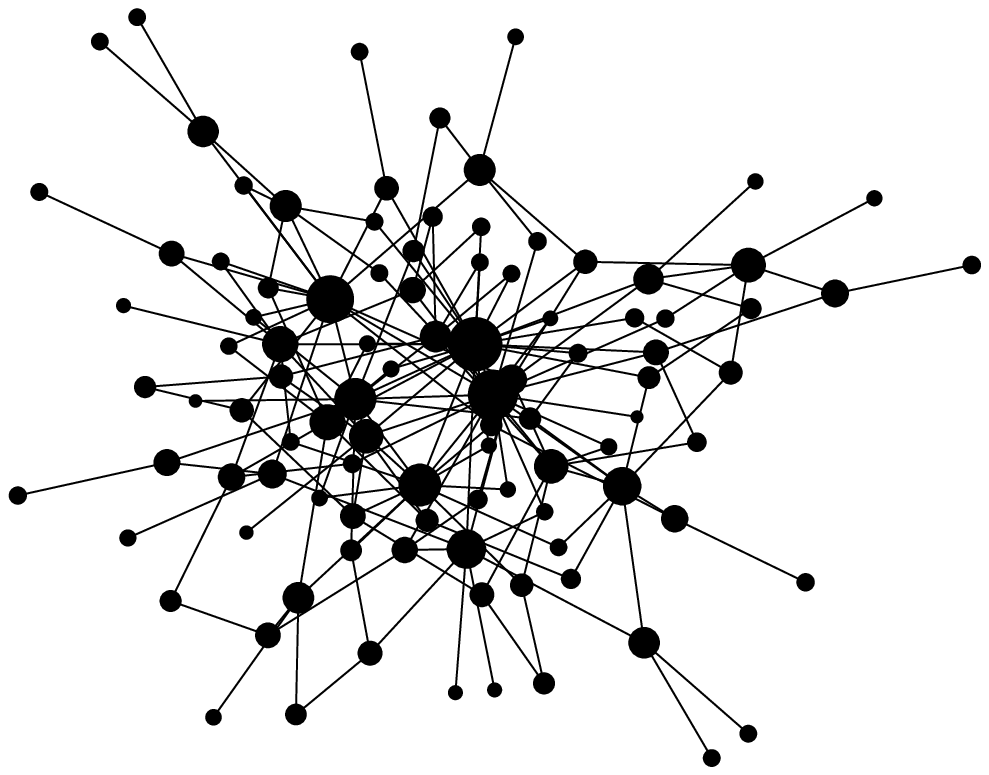}}
	\subfigure{\includegraphics[width=4.4cm]{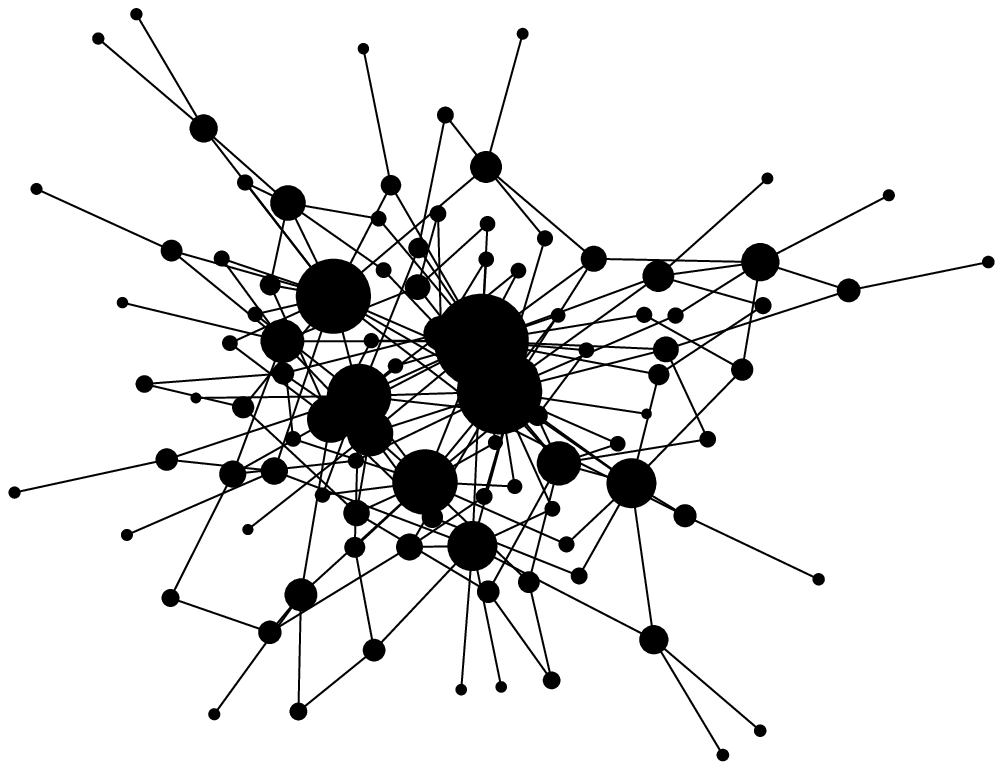}}
	\subfigure{\includegraphics[width=4.4cm]{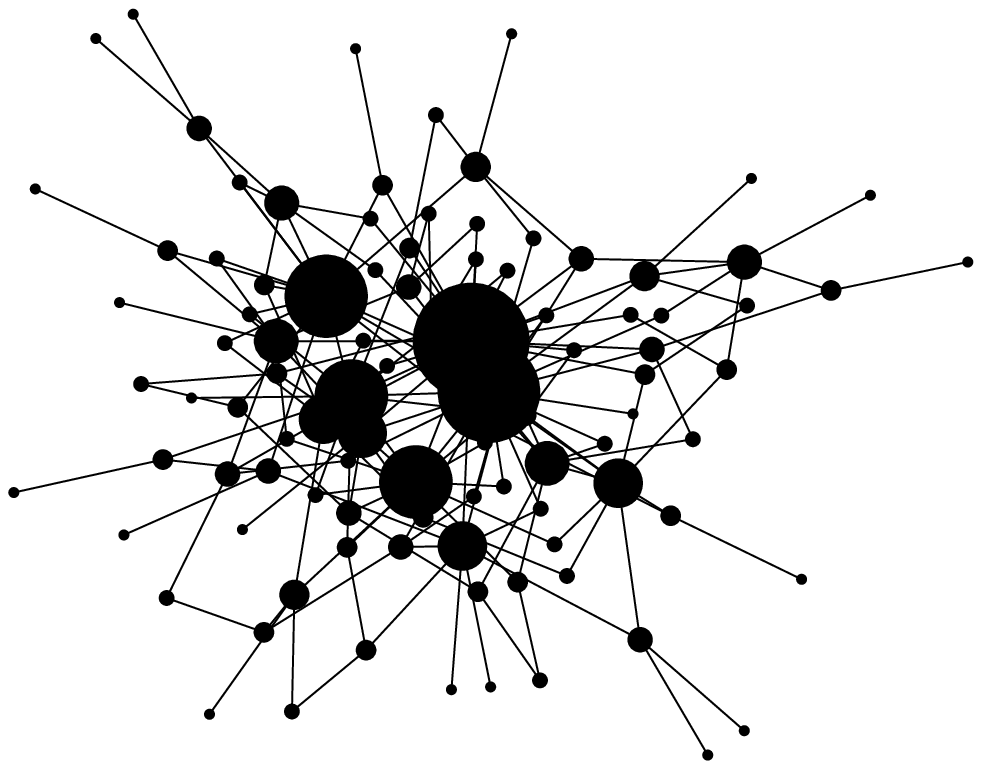}}
	\subfigure{\includegraphics[width=4.4cm]{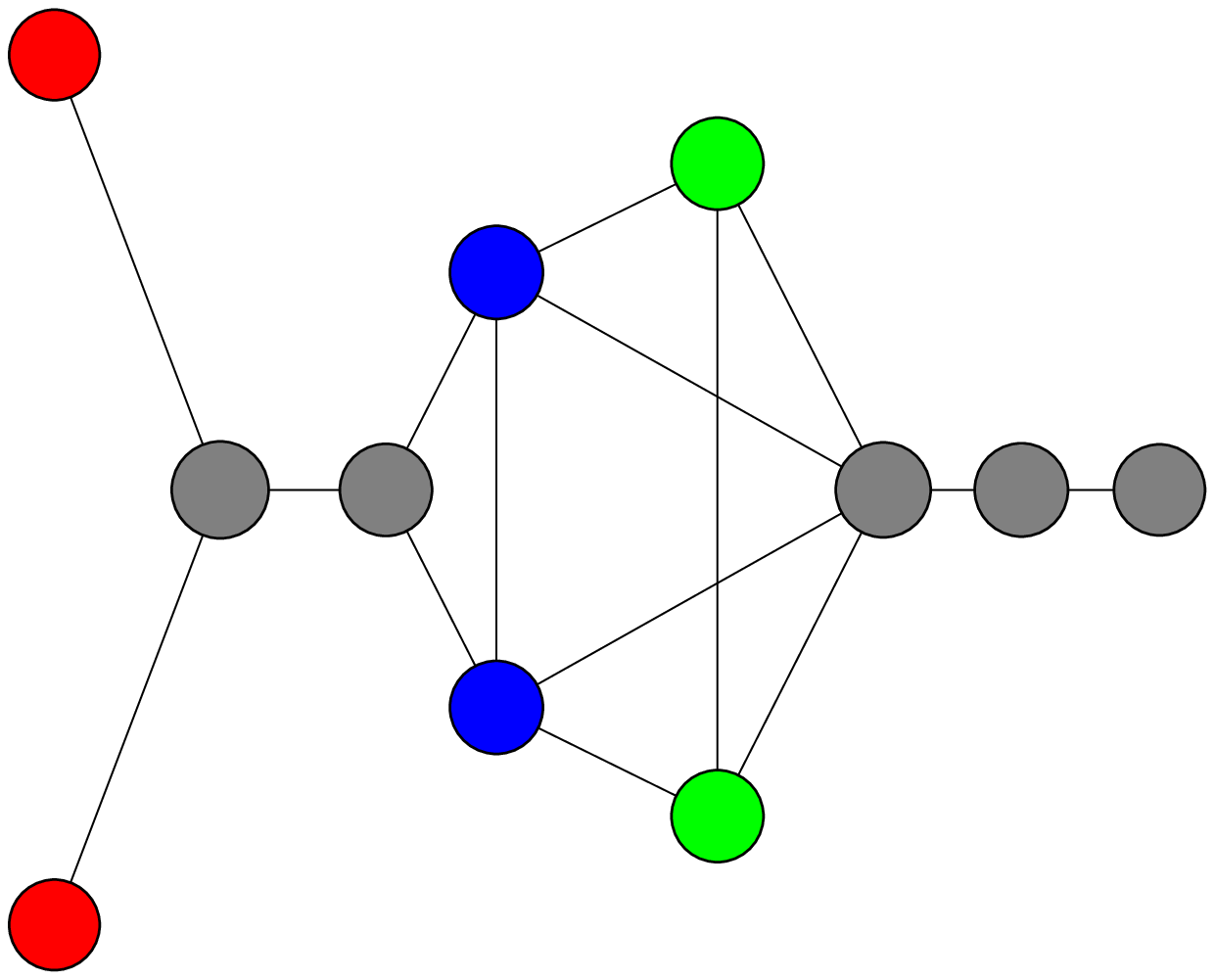}}
	\subfigure{\includegraphics[width=4.4cm]{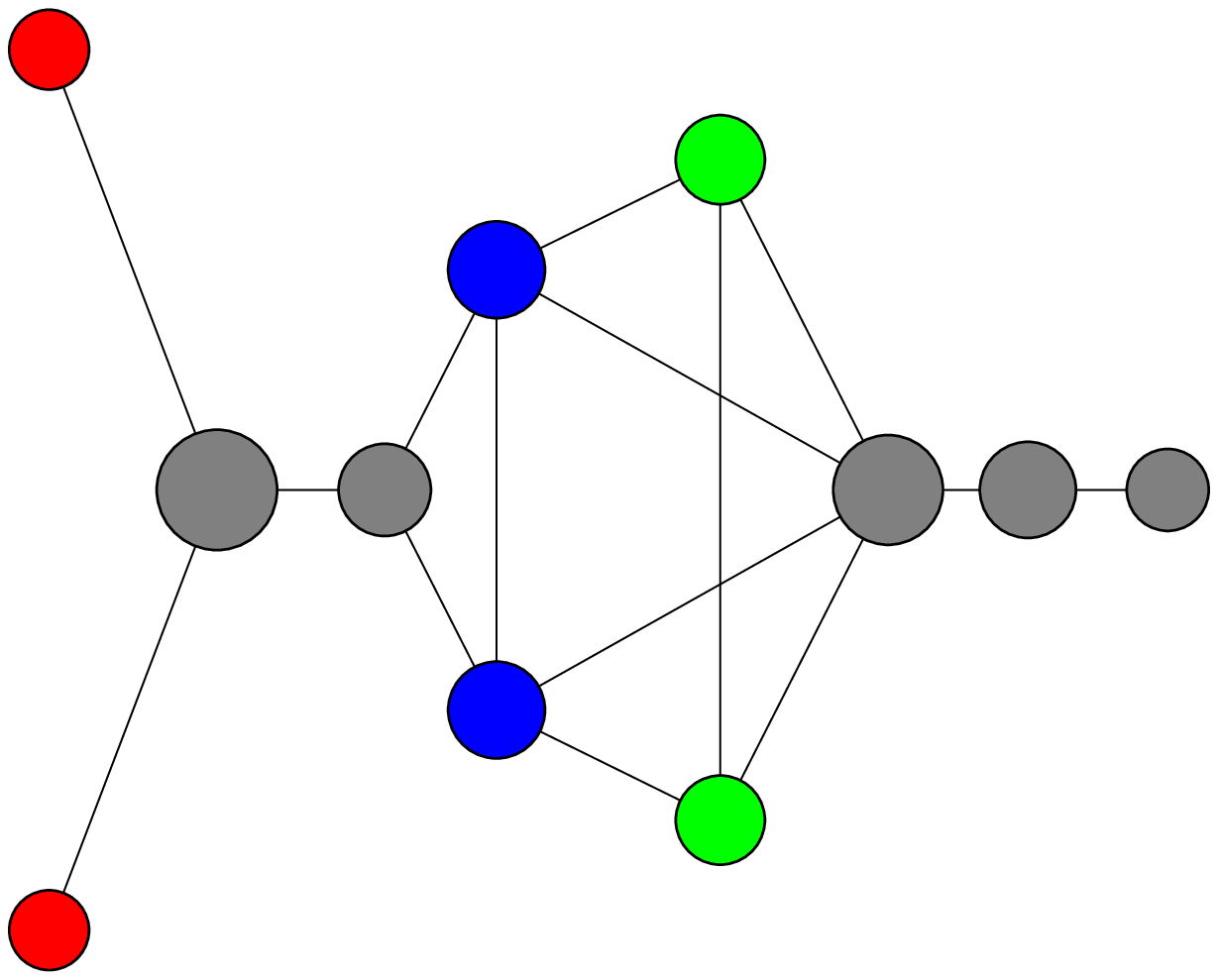}}
	\subfigure{\includegraphics[width=4.4cm]{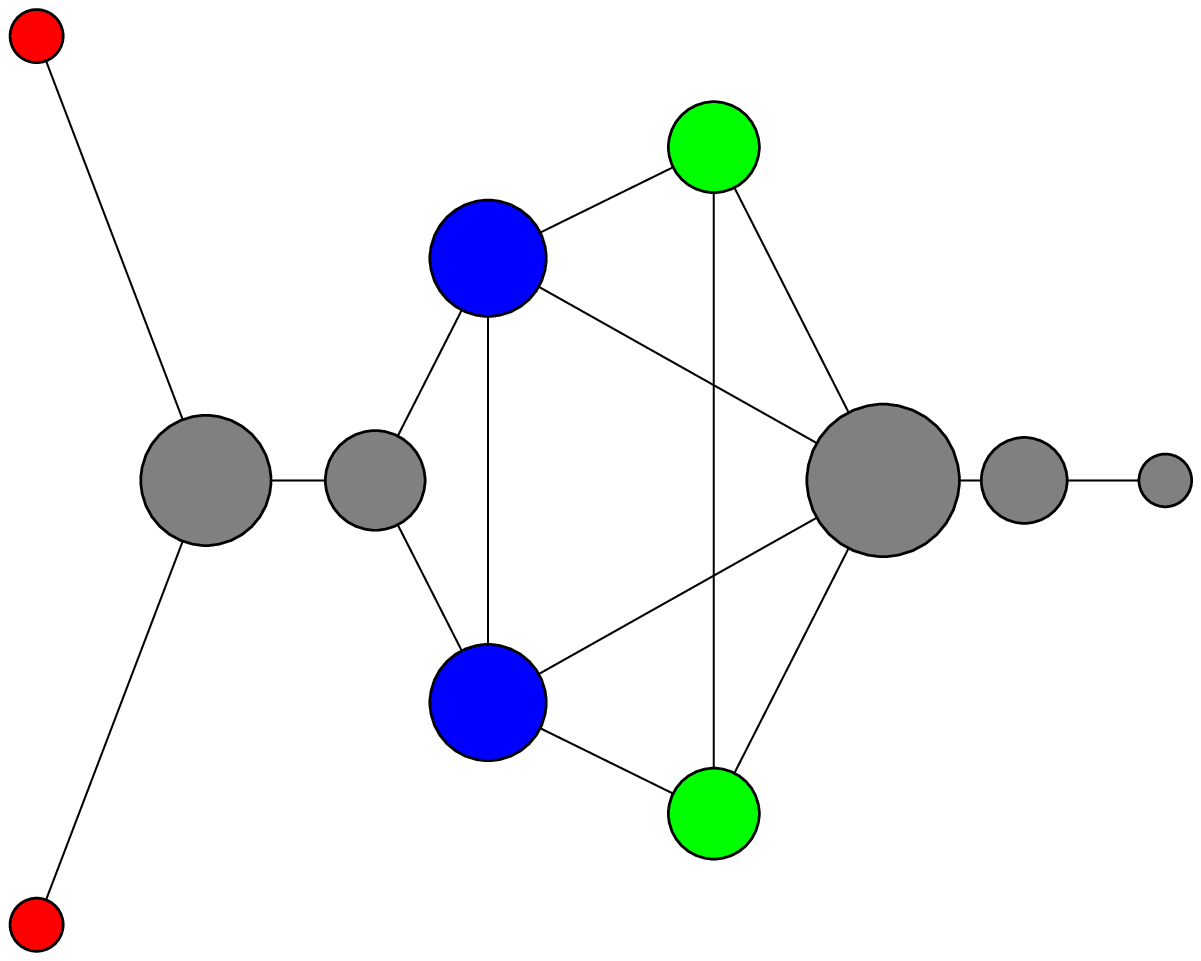}}
	\subfigure{\includegraphics[width=4.4cm]{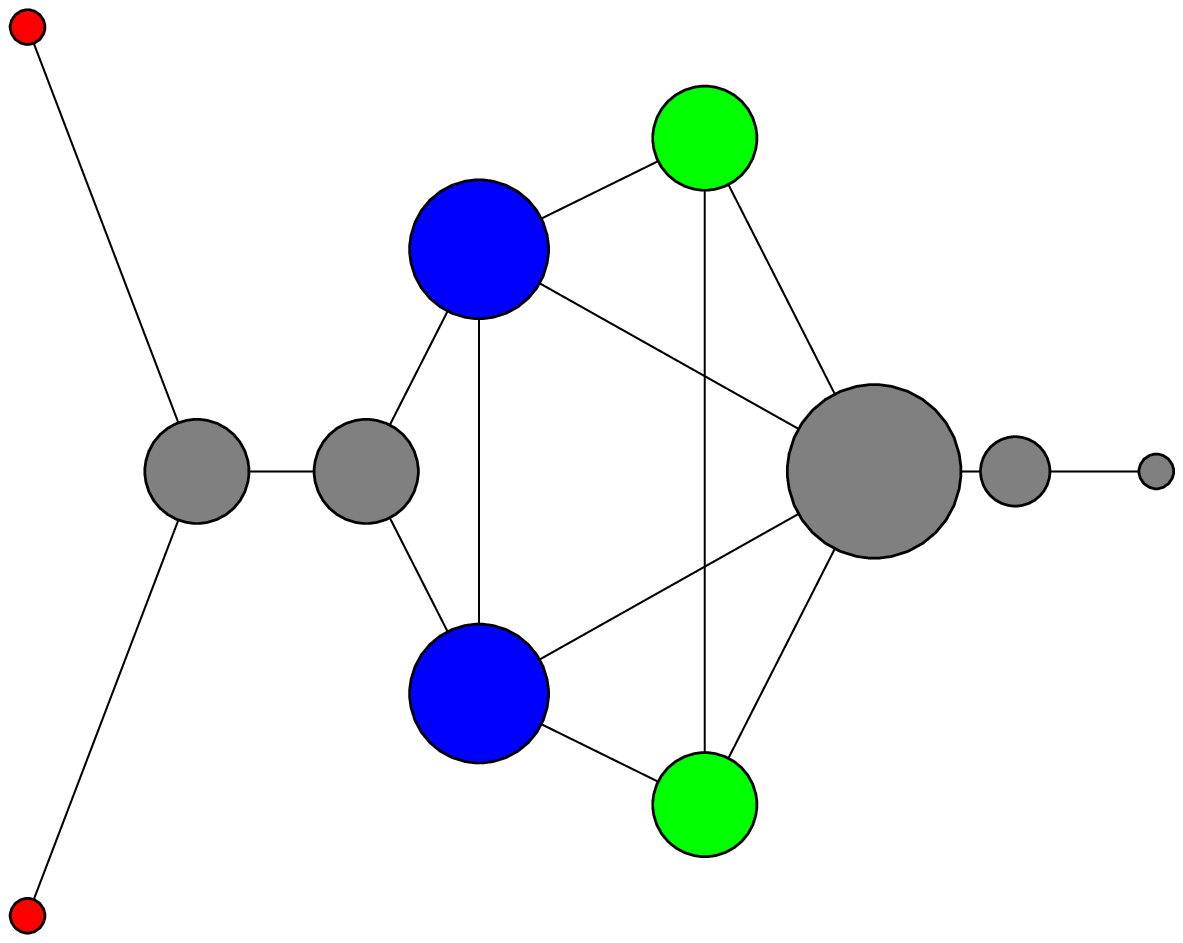}}
	\stackunder[5pt]{\includegraphics[width=4.3cm]{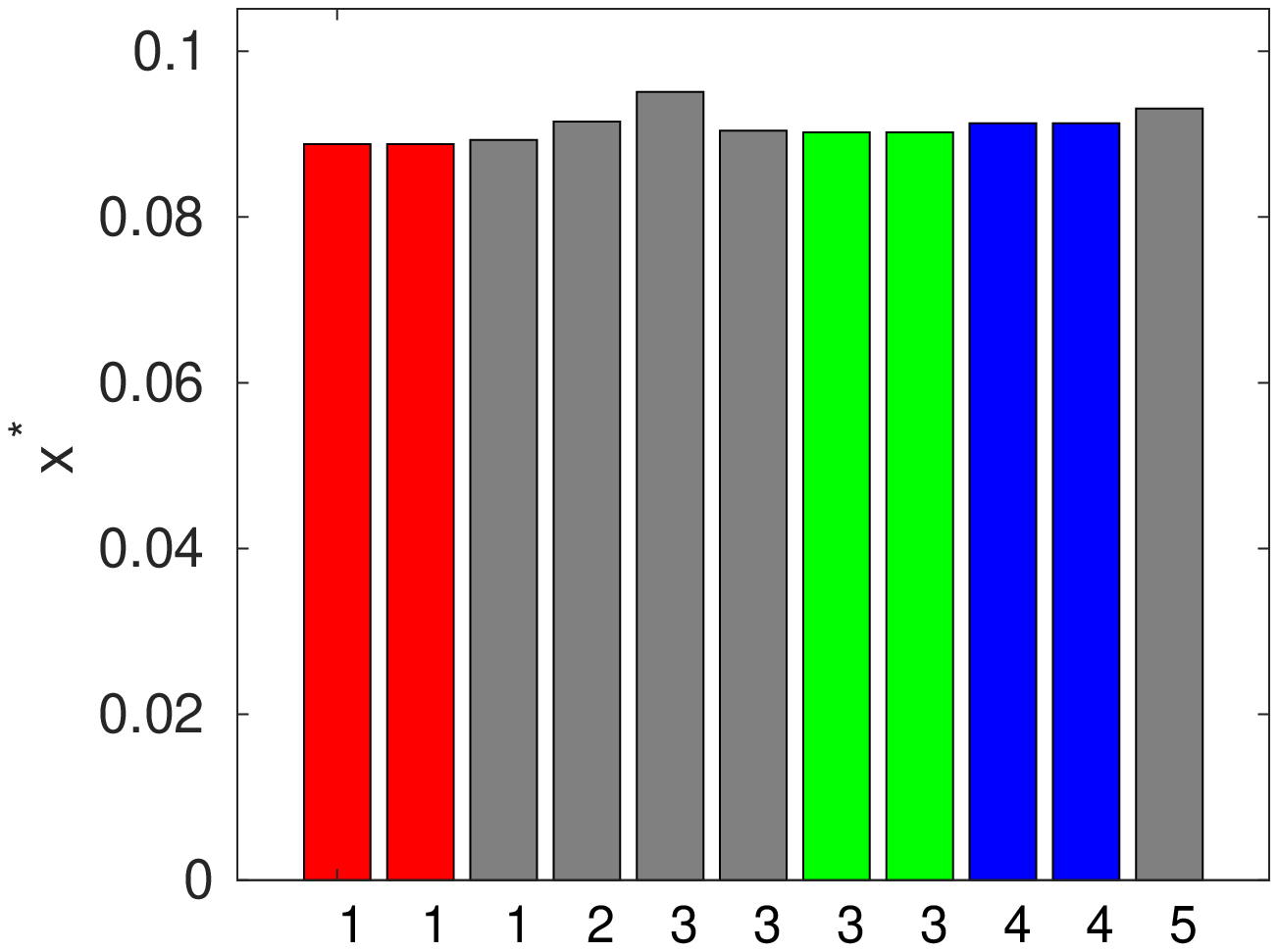}}{$\mu=0.1$}
	\stackunder[5pt]{\includegraphics[width=4.3cm]{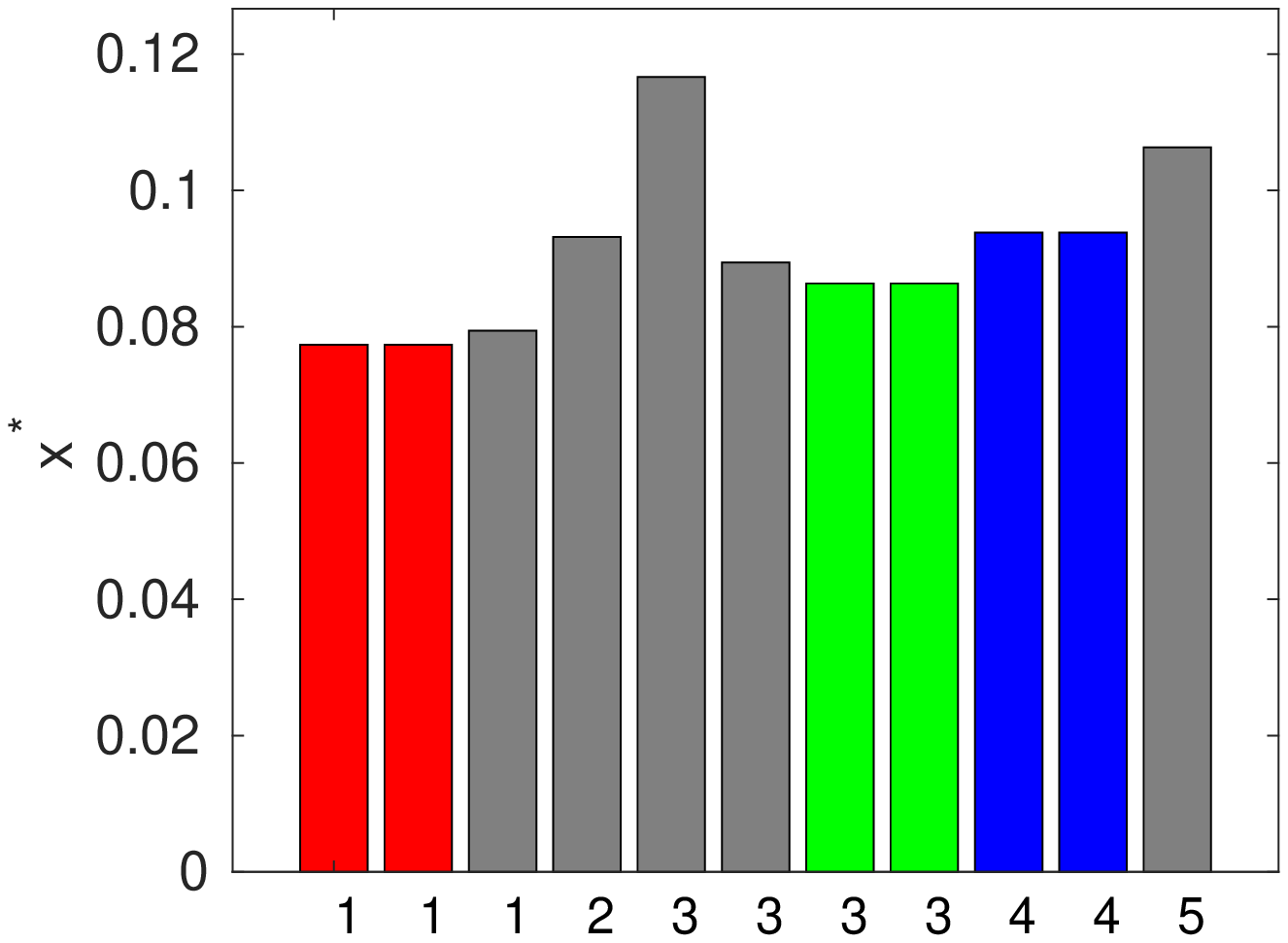}}{$\mu=0.5$}
	\stackunder[5pt]{\includegraphics[width=4.3cm]{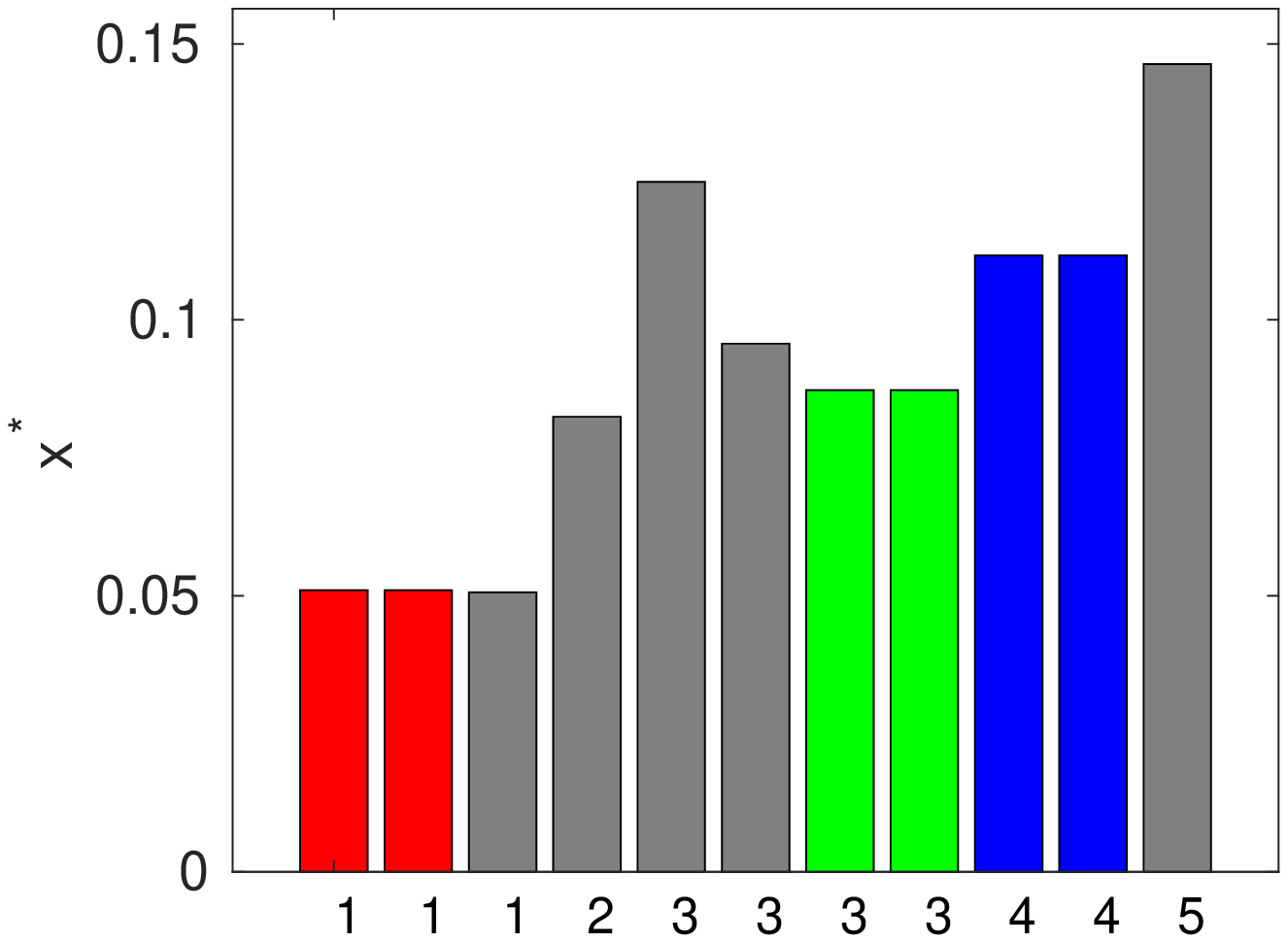}}{$\mu=0.9$}
	\stackunder[5pt]{\includegraphics[width=4.3cm]{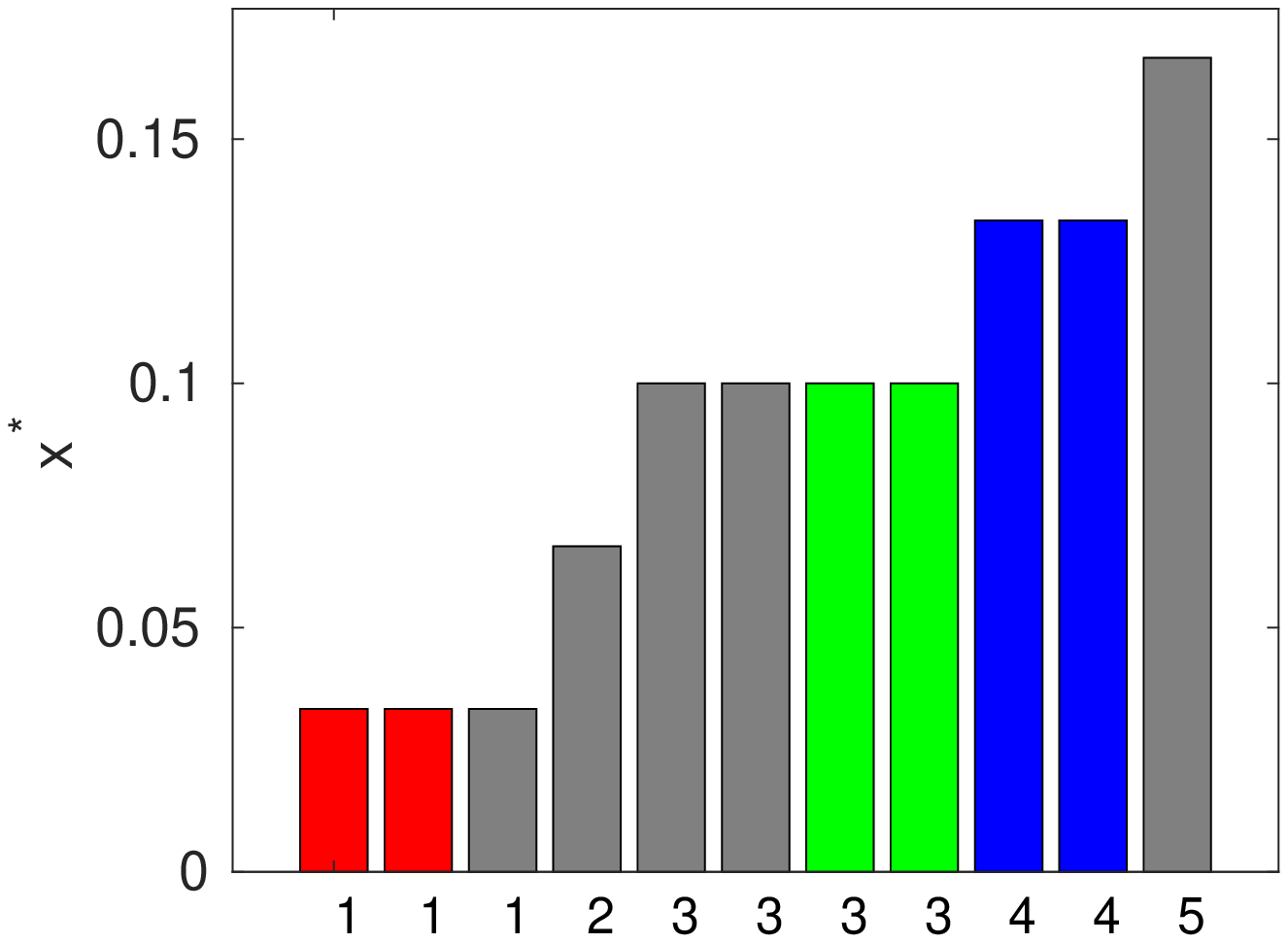}}{$\mu=1$}
	
	\caption{Reactive random walks on a scale-free network with $p_k\simeq
		k^{-\gamma}$ and $\gamma=2.2$, $N=100$ and mean degree $\langle
		k\rangle=3.6$, and on a toy graph with only 11 nodes. The four
		columns represent four different values of the mobility parameter:
		$\mu=0.1$, $\mu=0.5$, $\mu=0.9$ and $\mu=1$, where the size of the
		nodes is proportional to the different components of 
		$\boldsymbol{x^*}$. The reaction function selected are 
		$f(x)=x-x^2$ for the first network, and $f(x)=\sin(3x)$
		for the second one. In the second graph, the same colour has been
		used for nodes with the same symmetry, while colour gray has been used
		for all other (non-symmetric) nodes. The histograms
		report the stationary occupation at each node, while
		the node degree is indicated on the $x$ axis.}
	\label{fig:graphs}
	\vspace{1mm}
\end{figure}

%
Our third and last numerical example is reported in
Fig. \ref{fig:graphs}.  In this case, we have considered two different
topologies, namely a scale-free network with $N=100$ nodes (first row
panels) and a smaller network with $N=11$ nodes (second row
panels). Again, the stationary occupation probability at the nodes of
the graphs is shown for various values of $\mu$.  For both networks,
the size of the nodes in the graphs is proportional to
$\boldsymbol{x^*}$, while the four different columns represent respectively 
the four values of the mobility parameter, $\mu= 0.1, 0.5, 0.9, 1$.
While all the nodes have almost equal size for small
values of $\mu$, they clearly tend to differentiate when $\mu$ 
increases. Notice that for $\mu=1$ the node size only reflects  
their degree, so that the nodes with the largest sizes are the hubs of the
scale-free network in the first row. 
For intermediate values of the mobility parameter (see
for instance $\mu=0.9$), instead the nodes with the largest occupation
probability are those connecting isolated vertices to the rest of the
network, irrespective of their own
degree. This is evident for the second graph in the second and third
rows. For this graph symmetric nodes are also highlighted
in figure (see Section
\ref{sec_symm} for a formal definition of symmetric nodes) by adopting
the same colours for pairs of nodes with the same symmetry, and reporting
in gray nodes not having a symmetric counterpart.

\onecolumngrid

%

%
\newpage
\twocolumngrid


\section{Analytical derivation of the stationary state}
\label{sec_analytic}

\begin{figure}
	\subfigure[]{\includegraphics[width=8cm]{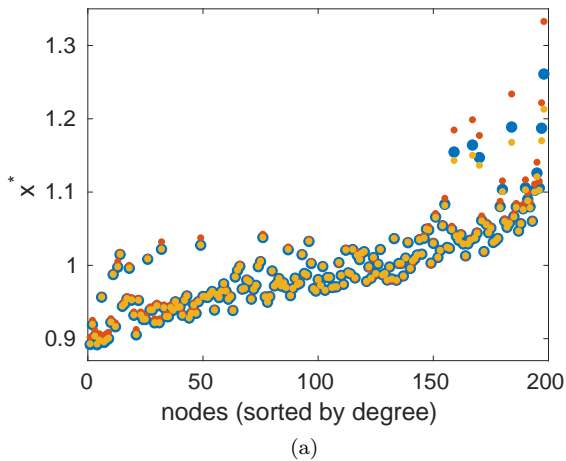}}
\caption{Comparison between analytical predictions (first order in red
  and second order in yellow) and numerical results (blues dots). The
  stationary occupation probability of different nodes (sorted by
  their degree) is shown for reactive random walkers with logistic
  growth $f(x)=x-x^2$ on graph of collaboration among jazz
  musicians~\cite{jazz_net}.  The mobility parameter $\mu$ has been
  set to 0.1.}
	\label{fig_anal_num}
\end{figure}

The fixed point $\boldsymbol x^*$ of the reactive random walk model 
in Eqs. \eqref{gamma_dynRW} is in
general
not easy to obtain
analytically because of the interplay between random walk dynamics and 
local interactions. Approximate techniques can be however employed in the
low-mobility limit $\mu\simeq0$, when the local dynamics is only
slightly modified by coupling between network nodes due to the movement.
In this limit, it is
possible to derive a perturbative estimate for $\boldsymbol x^{*}$:
$\boldsymbol x^* = s^*\boldsymbol1 +
\sum_{n=1}^{\infty}\mu^n \boldsymbol{\delta x}^{(n)}$, where
$\boldsymbol{\delta x}^{(n)}$ stands for the $n$-th correction to the
uncoupled case. The first two corrections take the explicit form:

\begin{equation}
\delta x_i^{(1)} = -\frac{s^*}{f'(s^*)}\sum_j l^{\rm RW}_{ij}
\label{deltax1}
\end{equation}
and
\begin{equation}
\begin{split}
\delta x_i^{(2)} &= -\frac{(s^*)^2}{2}\frac{f''(s^*)}{f'(s^*)^3}\biggl(\sum_j l^{\rm RW}_{ij}\biggr)^2\\ 
&-\frac{s^*}{f'(s^*)} \sum_j l^{\rm RW}_{ij} + \frac{s^*}{f'(s^*)^2}
\sum_j l^{\rm RW}_{ij}\sum_k l^{\rm RW}_{jk}=\\
&= \delta x^{(1)}_i - \frac{f''(s^*)}{2f'(s^*)}(\delta x^{(1)}_i)^2 - \frac{1}{f'(s^*)} \sum_j l^{\rm RW}_{ij}\delta x^{(1)}_j
\end{split}
\label{deltax2}
\end{equation}
where $s^*$ is the solution for $\mu=0$, $f(s^*)=0$.
In figure \ref{fig_anal_num} we show
that the analytical predictions are in agreement with the numerical solution.
In particular, we consider reactive random walkers with a mobility parameter
$\mu=0.1$ and a logistic function as local interaction term, 
and we implement the model on the graph of collaborations among jazz
  musicians~\cite{jazz_net}. 
\\ 
If $f$ is a $C^{\infty}$ function, the perturbative terms can be
computed for each order $n$. In this case the hypothesis of small
$\mu$ can be relaxed and the analytical solution for the fixed
point can be, in principle, exactly determined. In such a case, the generic $n$-th
correction can be cast in the form:

\onecolumngrid
\vspace{10pt}
\vspace{6pt}


\begin{equation}
\begin{split}
\delta x_i^{(n)} =& - \frac{1}{f'(s^*)} \Biggl\{\sum_{r=2}^{n} \frac{f^{(r)}}{r!} \Biggl[ \sum_{m_1=1}^{n-r+1} \hspace{2mm} \sum_{m_2=1}^{n-r-m_1+2}\hspace{2mm} \sum_{m_3=1}^{n-r-m_1-m_2+3}...\sum_{m_{r-1}=1}^{n-\sum_{j=1}^{r-2}m_j-1} \delta x_i^{(m_1)} \delta x_i^{(m_2)}...\delta x_i^{(m_{r-1})}\delta x_i^{(n-\sum_{k=1}^{r-1}m_k)} \Biggr] \\
& - \sum_{r=2}^{n-1} \frac{f^{(r)}}{r!} \Biggl[ \sum_{m_1=1}^{n-r} \hspace{2mm} \sum_{m_2=1}^{n-r-m_1+1} \hspace{2mm} \sum_{m_3=1}^{n-r-m_1-m_2+2}...\sum_{m_{r-1}=1}^{n-\sum_{j=1}^{r-2}m_j-2} \delta x_i^{(m_1)} \delta x_i^{(m_2)}...\delta x_i^{(m_{r-1})}\delta x_i^{(n-1-\sum_{k=1}^{r-1}m_k)} \Biggr] \\
& - f'(s^*)\delta x_i^{(n-1)} + \sum_j l^{\rm RW}_{ij}\delta x_j^{(n-1)} 	\Biggr\} 
\end{split}
\label{deltaxn}
\end{equation}

\vspace{5pt}
\vspace{6pt}
\twocolumngrid

where $f^{(r)}$ is the $r$-th derivative computed in $s^*$.\\
As expected, at different perturbative orders the local dynamics involves successive derivatives of $f$ at $s^*$. In particular, the first correction $\boldsymbol{\delta x}^{(1)}$ is only sensitive to the the first derivative, while in $\boldsymbol{\delta x}^{(2)}$ the second derivative appear. In general, the $n$-th correction is characterized by all the derivatives of $f$ until the $n$-th one. 
More interestingly it is worth noticing that $\delta x^{(1)}_i$  contains a term that, when the random walk is unbiased, is proportional to $\sum_j a_{ij}/k_j$, which essentially is a sum over all neighbours of node $i$ of their inverse degree. This implies that the first correction to the generic $i$-th component of the uniform fixed point depends on the inverse degree of all the nodes of the graph that are adjacent to $i$. In the second order correction, we instead find the term $\sum_{jl} a_{ij}/k_j a_{jl}/k_l$. The fixed point computed at the second order in $\mu$ thus not only depends on the inverse degree of the nearest neighbours of node $i$, but also on the inverse degree of its second-nearest neighbours. By iterating forward this reasoning, the $n$-th correction will depend on the $n$-th nearest neighbours degrees: the term $\sum_j l^{\rm RW}_{ij}\delta x_j^{(n-1)}$ in eq. \eqref{deltaxn}  takes recursively into account all the nodes of the network that  can be reached, in at most $n$ time steps, when starting from node $i$. Obviously, when $n$ goes to infinity all the nodes of the network contribute with their inverse degree.

It is also worth observing that the perturbative calculation can be
readily extended to the general case of biased random walks. To this
end one should consider the more general Laplacian form in the last
term of each correction: $\sum_j l^{\rm RW}_{ij}\delta x_j^{(n-1)}$.
In this case, the first correction $\delta x_i^{(1)}$ is not solely
influenced by first neighbours of node $i$, but also depends on the
second neighbours, being proportional to $k_i^{\alpha}\sum_j\frac{
  a_{ij}}{\sum_l a_{lj}k_l^{\alpha}}$.  Analogously, for the second
correction term, the biased random walks introduces a dependency on
the neighbours of all nodes at distance two from each vertex, and so
on. In general, considering a degree bias always has the effect of
moving the set of involved nodes to further proximity
level in the network, as already observed in~\cite{GomezGardenesLatora08} in the
case of non-reactive random walks.

In the next three sections we will explore how the occupation
probability of reactive random walkers can turn useful to define novel
measures of functional centrality for the nodes of a network,
to detect network symmetries, or to distinguish assortative from
disassortative networks.

\section{Measures of functional ranking}
\label{sec_centr}

Centrality measures allow to rank the nodes according to their
location in the network~\cite{LatoraNicosiaRusso17book}. Originally
employed in social network analysis to infer the influent actors in a
social system, but soon adopted in many other fields, different
centrality measures have been constructed to capture different aspects
which make a node important, from the number and strength of its
connections to its reachability. Commonly used centrality measures are
the eigenvector centrality~\cite{Perron07, Frobenius12}, the
$\alpha$-centrality~\cite{Bonacich72,BonacichLloyd01}, the betweenness
centrality~\cite{Freeman77}, the closeness centrality~\cite{Freeman78}
and, of course the simplest one, the degree centrality.
This latter corresponds to the fixed point of our model in the limit
$\mu=1$. In this case, the stationary occupation probability $x^*_i$
is indeeed proportional to the degree of node $i$. However, in our
model of reactive random walkers, when $\mu \neq 1$, the stationary state
of the model will also depend on the choice of the local dynamics,
resulting in a
plethora of distinct configurations fostering different roles within
the network. In other words, for a fixed value of the mobility
parameter we can interpret our dynamical system as a
\textit{reaction-dependent centrality measure}. Moreover, we note that
the form of Eq. \eqref{conv_eq} on which this centrality measure is
based, is reminiscent of other existing definitions of
centralities such as a generalization of the
Bonacich centrality~\cite{Bonacich72} 
known as the $\alpha$-centrality ~\cite{BonacichLloyd01}, 
and the PageRank centrality
(PRC)~\cite{BrinPage98}. For instance, the PageRank centrality $x_i^{\rm PR}$
of a graph node $i$ is defined as
~\cite{Langville04,Gleich15,Bryan06}:  
\begin{equation}
x_i^{\rm PR} = d \sum_j \frac{a_{ij}}{k_j}x_j^{\rm PR} + \frac{1-d}{N} 
\label{pageRank}
\end{equation}
where $d \in (0,1)$ is a parameter usually set equal to 0.85.  PRC was
originally proposed as a method to rank the pages of the World Wide
Web. Indeed, it mimics the process of a typical user navigating
through the World Wide Web as a special random walk with
``teleportation'' on the corresponding graph. Such a random walker
with a probability $d$ performs local moves on the graph (most of the
times a user surfing on the Web randomly click one of the links in the
page that is currently being visited), while with a probability $1-d$
starts again the process at a node randomly chosen from the $N$ nodes
of the graph (the surfer starts again from a new Web site).  The
latter action, the so-called ``teleportation'' is represented by the
term $(1-d)/N$ in Eq. \eqref{pageRank}.  Notice that the value of $d =
0.85$ is estimated from the average frequency at which surfers recur
to their browser's bookmark feature. The introduction of the teleportation
term assigns a uniform non-zero weight to each vertex, and it is particularly
useful to avoid pathological cases of nodes with null centrality, in
the case the graph is not connected (or strongly connected if a
directed graph). In some cases however the teleportation
contribution is not uniform, but can be designed to gauge an
intrinsinc importance of each node. 
This implies enforcing a dependence on the generic node index $i$ in
the second term in the right hand side of Eq. \eqref{pageRank}.  The
advantage of using Eq. \eqref{conv_eq} instead of Eq. \eqref{pageRank}
as a measure of centrality then consists in the possibility of freely
choosing the reaction term. 
The adoption of function $f(x_i)$ in Eq. \eqref{conv_eq},
assigning a different contribute
to each node $i$ that depends on $x_i$, finds a plausible justification 
in the fact that the importance of a node may also depend on other
factors, not necessarily directly linked to the topology of the graph,
such as the status or functionality of the node. In a social network,
for instance, this factor could be related to the age, social status
or income of an individual. Moreover, $f$ can be chosen so as to take
into account the temporal evolution of some features of the nodes of
the network. Let us consider again the problem of ranking Web pages.
PageRank centrality in  Eq. \eqref{pageRank} can be
modified by replacing the constant
teleportation term with a variable contribution, due for instance to
the number of visualizations of each page, which could be suitably
described by a non-constant term proportional to $x_i(t)$, or more
generally by a function $f(x_i)$ as in Eq. \eqref{conv_eq}. 

\begin{figure}[h]
	\subfigure[]{\includegraphics[width=8cm]{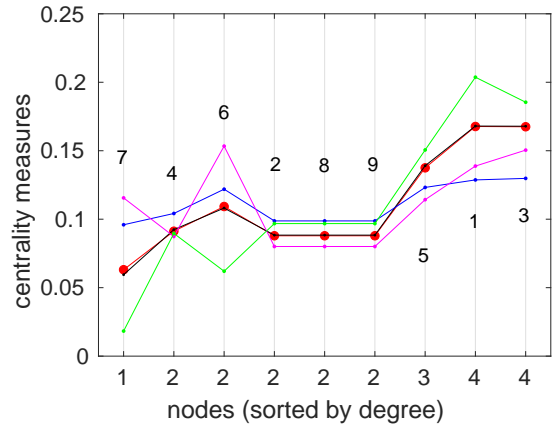}
		\label{fig_centr_a}}
	\subfigure[]{\includegraphics[width=8cm]{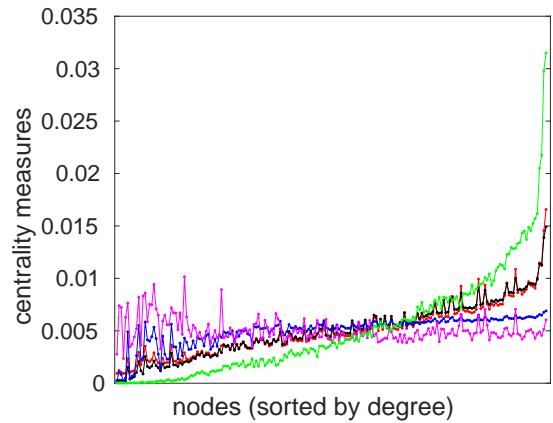}
		\label{fig_centr_b}}
	\caption{Measures of centrality based on Eq. \eqref{conv_eq} and on
		different choices of $f$ and $\alpha$ are compared to PRC (red
		curves) in the case of two networks, the graph of $N=9$ nodes in
		Fig. \ref{fig_net9nodes} and the graph of $N=198$ nodes representing the jazz musicians network~\cite{jazz_net}.}
	\label{fig_centr}
\end{figure}

As a practical example let us come back to examining the graph in 
Fig. \ref{fig_net9nodes} and focus again on node 6. 
According to stardard centrality measures such a node would not
result as a very central one, being in
a peripheral part of the graph and having just two neighbours. 
However, one the neighbours is node 7, which is a graph leaf 
and this makes node 6 its only bridge towards the rest of the graph.
This consideration highlights the importance of nodes bridging other nodes
of the network and, depending on which characteristics we want to focus on,
could be an extremely useful feature to take into account when 
devising a measure of node centrality. 
Increasing the importance of this class of nodes can be for instance
obtained by an appropriate choice of function $f(x)$ in Eq. \eqref{conv_eq}. 
This is clearly shown in Fig. \ref{fig_centr_a}, where the rankings of
the graph nodes obtained for different reference reaction functions
and also for different choices of the mobility and bias parameters are
compared. The nodes are sorted according to their degree, which is
explicitly indicated on the x-axis, while the other reported
numbers correspond to node labels as in Fig. \ref{fig_net9nodes}.
Node 6, which bridges node 7 to the rest of the graph, appears to be more
sensitive than the others to the changes, with  
a large variety of ranking positions, especially if compared to the other
nodes with the same degree.  The magenta and green symbols
respectively refer to a positive ($\alpha=1$) and a negative ($\alpha=-1$)
bias with $f(x)=x-x^2$ and $\mu=0.85$. We observe that it is also
possible to reproduce the same trend of the PRC (red symbols) by again
using the logistic function with the same value of the mobility
parameter, but setting the bias to zero (black symbols). A different
reaction is used for the blue curve: $f(x)=x-x^{10}$ with $\mu = 0.7$
and $\alpha=0$.
\\
The same types of functional ranking as in Fig. \ref{fig_centr_a} have also
been adopted in Fig. \ref{fig_centr_b} for the nodes of the network
of collaborations among jazz musicians, and the results are 
reported with the same color code. A similar general trend appears,  
with low degree nodes enhanced by a negative degree bias and vice versa
hubs enhanced by a positive bias. In addition to this, we observe some 
fluctuations with peaks appearing in the different ranking measures,
most of them corresponding to nodes bridging one or more otherwise isolated
nodes of the network.  

In conclusion, the proposed measure of functional ranking 
can mimic other centrality measures, like PRC, in the limit of
large $\mu$, where diffusion is important and it is only slightly
modified by the local interactions. In 
general, for every value of $\mu$ between 0 and 1, our model of reactive
random walkers can be thought as a new way to measure centrality which
accounts for the differences between nodes at a deeper level, with 
the focus on different time-varying characteristics of
the nodes themselves.

\section{Detecting network symmetries}
\label{sec_symm}

Symmetries are ubiquitous in nature, and one of the main reasons by
which humans have been long attempted to describe and model the world
through the tools and the language of mathematics. In complex
networks, despite the fact that symmetric nodes may appear as special
cases, they are surprisingly numerous in real and artificial network
structures~\cite{SiddiquePecoraSorrentino17}.

In mathematical terms, network symmetries form a group, each element
of which can be described by a permutation matrix that re-orders
the nodes in a way that leaves the graph unchanged. More precisely, 
a graph $G$ with $N$ nodes described by the adjacency matrix $A$ 
has a symmetry if there exists a permutation matrix $P$,
i.e. a $N \times N$ matrix with each row and each column
having exactly one entry equal to 1 and all others 0, such
that $P$ commutes with $A$: $PA=AP$. 
This is equivalent to say that $PAP^{-1}=A$, namely that 
$PAP^{-1}$ performs a relabeling of the nodes of the original
graph which preserves the adjacency matrix $A$. 
Therefore, two nodes of the graph are said \textit{symmetric} if
their swapping preserves the adjacency relation. 
This implies that two symmetric nodes
are necessarily characterized by the same degree, but also that their
neighbours must have the same degree, so as the neighbours of their
neighbours, and so on.

While network symmetries may be easy to spot in small graphs like
those considered in Fig. \ref{fig_9nodes} and in
Fig. \ref{fig:graphs}, this is typically not the case for large
graphs. Different techniques to reveal symmetries in networks have
been developed, both numerical and analytical~\cite{Nicosia_etal13,PecoraSorrentino_etal14, SiddiquePecoraSorrentino17,ZhangMotterNishikawa17}. As we will
show below, reactive random walkers provide another method to
detect symmetric nodes by looking at the value of 
the stationary occupation probability at different nodes. 
In fact, while in the case $\mu=1$ of a pure random walk
process the fixed point $\boldsymbol x^*$ is solely determined by
the node degrees, when  $\mu \neq 1$ the dynamics is governed by
the network as a whole and the value of the
stationary occupation probability at a node will depend of its
degree, but also on the properties of the second, third and so on
neighbours. Hence, it is plausible to conclude that only perfectly
symmetric nodes
can assume the same asymptotic occupation probability, and to propose
to detect symmetric nodes of a graph by looking at those having the same
value of $\boldsymbol x^*$ for a reactive random walker model with $\mu \neq 1$
on the graph.

An analytical argument in support of this
can be obtained from the perturbative derivation of
the stationary state presented in Section \ref{sec_analytic}. In the
limit $\mu\simeq0$, the expression for the first correction
$\delta x^{(1)}_i$ to the uniform stationary state given in 
Eq. \eqref{deltax1} contains a term proportional to
$\sum_j a_{ij}/k_j$, which indicates the dependence of the stationary
state on the degree of the neighbours of $i$. Analogously, 
the degree of the second nearest neighbours can be found in the 
second correction $\delta x^{(2)}_i$ , while the degree of the
$n$-th nearest neighbours appears in the $n$-th correction.
The value of $x^*_i$ of a node $i$ will consequently depend on the degrees of
all the nodes in the graph. Since two symmetric nodes share the
same connectivity at each level of neighbourhood, 
we can then find symmetric nodes as those with exactly the same value
of $x^*$.

Let us come back to the graphs considered in Figs. \ref{fig_9nodes}
and \ref{fig:graphs}. In the first example the three nodes
labeled as 2, 8 and 9 are symmetric, as can be seen directly from figure
\ref{fig_9nodes} (b), given that they share the same set of
neighbours. The existence of such a symmetry is also revealed
by looking at the behaviour of the occupation probability of
different nodes when varying $\mu$:  
Fig. \ref{fig_9nodes}(b) shows that the curves corresponding to these
three nodes are indistinguishable.
Another remarkable example is 
reported in Fig. \ref{fig:graphs}, where the graph reported 
in the second row panels is taken as reference
model to observe the variation in the occupation probability state for
different values of $\mu$. Here, nodes with the same symmetries are
shown with the same colour, while the remaining nodes are in grey, and
correspond to exactly the same value of $\boldsymbol x$, as reported
in the third row panels of the same figure.

A more general argument that extends the results above from
$\mu\simeq0$ to the general case $\mu \neq 1$ can be obtained by
proving that Eqs. \eqref{gamma_dynRW} are equivariant under a
permutation of symmetric nodes \cite{PecoraSorrentino_etal14}.
Such equations can be rewritten in vectorial notation as: 
\begin{equation}
\boldsymbol{\dot{x}} = A K^{-1} \boldsymbol{x} -\boldsymbol{x}
+ \mathcal{F}(\boldsymbol{x})
\label{RWdyn_vec}
\end{equation}
where $K = \{ k_{ij} \}$ is a diagonal matrix whose entries are 
defined as $ k_{ij} = k_i \delta_{ij}$, 
and the functional $\mathcal{F}:\mathbb{R}^N \rightarrow \mathbb{R}^N$
is defined such that the generic $i$-th element of the image vector
$[\mathcal{F}(\boldsymbol{x})]_i$ is equal to $f(x_i)$. 
Our goal is now to prove
that Eq. \eqref{RWdyn_vec} also holds for the permuted
vector $P \boldsymbol{x}$. Left-multiplying the equation by
matrix $P$ we get:
\begin{eqnarray}
P \boldsymbol{\dot{x}} &=& P A K^{-1} \boldsymbol{x} - P
\boldsymbol{x} + P \mathcal{F} (\boldsymbol{x})=
\nonumber
\\
&=& A K^{-1} P \boldsymbol{x}  - P \boldsymbol{x} +P \mathcal{F} (\boldsymbol{x}) 
\label{RWdyn_vecP}
\end{eqnarray}
where in the last equality we have used
the fact that $P$ commutes with $A$ and, since symmetric nodes
have the same degree, it also commutes with $K$ and consequently with
its inverse.
Now we observe that the role of matrix $P$ is to permute symmetric
nodes leaving the others unchanged. The effect of $P$ on a generic
vector $\boldsymbol v \in \mathbb{R}^N$ is $[P
\boldsymbol{v}]_i=v_{\tilde i}$ where $\tilde i$ denotes the node of
the network which is the symmetric twin of $i$, if it exists, otherwise
$\tilde i=i$. Consequently, when we apply $P$ to 
$ \mathcal{F} (\boldsymbol{x})$ we obtain a vector whose $i$-th 
component is:
\begin{equation}
[ P  \mathcal{F}(\boldsymbol{x})]_i = [ \mathcal{F}(\boldsymbol{x})]_{\tilde i} =
f(x_{\tilde i}) = f([P \boldsymbol{x}]_i) = [\mathcal{F}( P\boldsymbol{x})]_i
\end{equation}
Making use of this result, Eq. \eqref{RWdyn_vecP} becomes the equivalent
of Eq. \eqref{RWdyn_vec} evaluated for $P \boldsymbol{x}$ instead of
$\boldsymbol{x}$, which is what we wanted to prove.


\section{Measuring degree correlations}
\label{sec_degreecorr}


A distinguishing feature of many real-world networks is the presence
of non-trivial patterns of degree-degree
correlations~~\cite{PastorSatorrasVazquezVespignani01,Newman02,Newman03}.
In the case of positive degree-degree correlation the network is said to be
\textit{assortative}: this is often the case for social networks,
where hubs have a pronounced tendency to be linked to each
other. Conversely, a network is said \textit{disassortative} if the
correlations are negative and connections between hubs and poorly
connected nodes are favored.
Well-known examples of disassortative
networks are the Internet, and biological networks such as
protein-protein interaction networks, where high degree nodes tend to
avoid each other.

One possible way to reveal the presence of degree-degree correlations
in a network is to compute the average degree of neighbours of nodes
of degree $k$, and
to look at how this quantity depends on the value of $k$. 
The average degree $k_{nn,i}$ of the neighbours of node $i$ is defined
as $k_{nn,i}= \frac{1}{k_i}\sum_j a_{ij}k_j$. 
To obtain the average degree of neighbours of nodes of degree $k$, we need
to average the quantity $k_{nn,i}$ over all nodes $i$ of degree $k$. 
Let us denote as $p_{k'|k}$ the
conditional probability\footnote{To construct the conditional probabilities $p_{k'|k}$ it is convenient
to define a matrix $E$ such that the entry $e_{kk'}$ is equal to
the number of edges between nodes of degree $k$ and nodes of degree
$k'$, for $k \neq k'$, while  $e_{kk'}$ is twice the number of links
connecting two nodes having both degree $k$.
The conditional probability $p_{k'|k}$ can be then expressed as
$p_{k'|k}=e_{kk'}/\sum_{k'}e_{kk'}$~\cite{LatoraNicosiaRusso17book}. By
definition such a probability satisfies the normalization condition
$\sum_{k'} p_{k'|k} = 1\ \forall k$.} that a link from a node of degree $k$ is connected
to a node of degree $k'$.  Now, 
by expressing the sum over nodes as a sum over degree classes,
the average degree of the nearest neighbours
of nodes with a given degree $k$ can be written as: 
\begin{equation*}
\langle k_{nn}\rangle_k = \sum_{k'} k' p_{k'|k}.
\end{equation*}
The function $ \langle k_{nn}\rangle_k$ is a good indicator of the presence 
of degree correlations in a network. In fact, the quantity $ \langle k_{nn}\rangle_k$ 
increases with $k$ when the network has positive degree correlations,
it is decreasing when the network has negative correlations, while it is constant and
equal to $\langle k^2\rangle/\langle k\rangle$ for uncorrelated networks.
\\
We will now show that the dynamics of the reactive random walker model
of Eq.~\eqref{gamma_dynRW} is sensitive to the presence of
correlations in the the underlying network, and it is therefore possible
to detect and measure the assortative or disassortative nature of
a network from the asymptotic node occupation probability.  
To this end we need to return to the perturbative approach
to obtain the equilibrium occupation probability discussed 
in Section \ref{sec_analytic}. As already remarked,
a full hierarchy of terms are found to appear as a byproduct of the
calculation, which respectively relate to paths connecting nodes that are $1, 2,
3,\ldots$ steps away from any selected node. 
Let us focus on the first correction to the uniform state, 
namely the term $\delta x^{(1)}_i$, as specified
in Eq.~\eqref{deltax1}. Up to the multiplicative node-invariant factor
$s^*/f'(s^*)$, $\delta x^{(1)}_i$ is equal to $\sum_j l^{\rm RW}_{ij}
= \sum_j {a_{ij}}/{k_j} - 1$. Therefore, at the first order, the difference
between the equilibrium distribution and the uniform state is governed by
the quantity:  
\begin{equation}
w_i^{(1)} \equiv \sum_{j=1}^N \frac{a_{ij}}{k_j} = {k_i}{ \langle \frac{1} {k_{nn}} \rangle   }  ,
\label{eq_wi1}
\end{equation}
representing, for a generic node $i$, the sum of the inverse degrees
of all its neighbours. The quantity $w_i^{(1)}$ is always
non-negative, and it gets larger when many nodes are adjacent to node
$i$ (large degree $k_i$, corresponding to many terms in the sum) and
all such nodes display smaller degrees.  In the particular case in
which all the nodes connected to $i$ have exactly degree equal to
$k_i$, we get $w_i^{(1)}=1$. When instead, the degree $k_i$ of node $i$ is
smaller than the inverse of the mean inverse degree of the nodes adjacent to $i$ , then we
have $0<w_i^{(1)}<1$. In the extreme case of low degree nodes
connected to hubs $w_i^{(1)}$ tends to zero\footnote{Using the
	definition given in section \ref{sec_centr}, the quantity
	$\boldsymbol{w}^{(1)}$ weights the role of $i$ in bridging the gap
	between neighbours. In other words, it gauges how much node $i$ is
	important in linking \textit{isolated nodes} to the main bulk, so
	keeping the graph connected. We already mentioned the role of node 6
	in the graph of Fig. \ref{fig_net9nodes} and how its intrinsic
	relevance stems from the stationary solution $\boldsymbol{x^*}$ (see
	Fig. \ref{fig_centr_a}).  The formal explanation of this
	phenomenon is indeed due to the presence of the quantity $\boldsymbol
	w$ in the first term of the perturbative expansion of
	$\boldsymbol x^*$.}.

Looking at the whole network, the vector $\boldsymbol w$
can be turned into an effective indicator for the presence of
degree-degree correlations that relies on the harmonic mean of the
degrees instead that on the standard mean.

For instance, we can consider the average value of $w_i^{(1)}$ for all
nodes $i$ of degree $k_i=k$. Such a quantity can be written in terms
of the adjacency matrix of the graph as: 
\begin{equation}
\langle w^{(1)} \rangle_k = \frac{1}{N_k}\sum_{i=1}^N \sum_{j=1}^N \frac{a_{ij}}{k_j} \delta_{k_i,k}
\end{equation}
where $N_k=\sum_{i=1}^N \delta_{k_i,k}$ is the number of nodes of
degree $k$.  We can rewrite the previous equation by making use of the
conditional probability $p_{k'|k}$, so that the sum over all
neighbours $j$ of $i$ becomes a sum over the degrees $k'$ of the nodes
adjacent to those of degree $k$. We finally obtain:
\begin{equation}
\langle w^{(1)} \rangle_k =  k\sum_{k'} \frac{1}{k'} p_{k'|k} = k \langle \frac{1}{k'}\rangle_k,
\label{eqwk} 
\end{equation}
where the quantity $\langle {1}/{k'} \rangle_k$ denotes the average of
the inverse degree of the first neighbours of nodes of degree $k$. 
In absence of degree correlations the conditional probability takes the form:
$p_{k'|k}^{\rm nc} = {k'p_{k'}} / {\langle k \rangle}$
~\cite{Newman02,LatoraNicosiaRusso17book}, where $p_k$ is the degree
distribution of the network, $\langle k \rangle$ is the average
degree, and ``nc'' stands for no correlations.  
Hence, in uncorrelated networks the quantity in Eq.~\eqref{eqwk}
reduces to: 
\begin{equation}
\langle w^{(1)}_{\rm nc} \rangle_k = \bigl( k\sum_{k'} \frac{1}{k'} p_{k'|k}^{\rm nc}\bigr) = \frac{k}{\avk}
\end{equation}
and is a linearly increasing function of $k$ with slope equal to
$1 / \avk $. 
Such a function represents the reference case to compare to 
when evaluating the quantity $\langle w^{(1)} \rangle_k$ for a
given network.  

\begin{figure}[H]
	\centering
	\subfigure[]{\includegraphics[width=10.1cm,height=7.9cm]{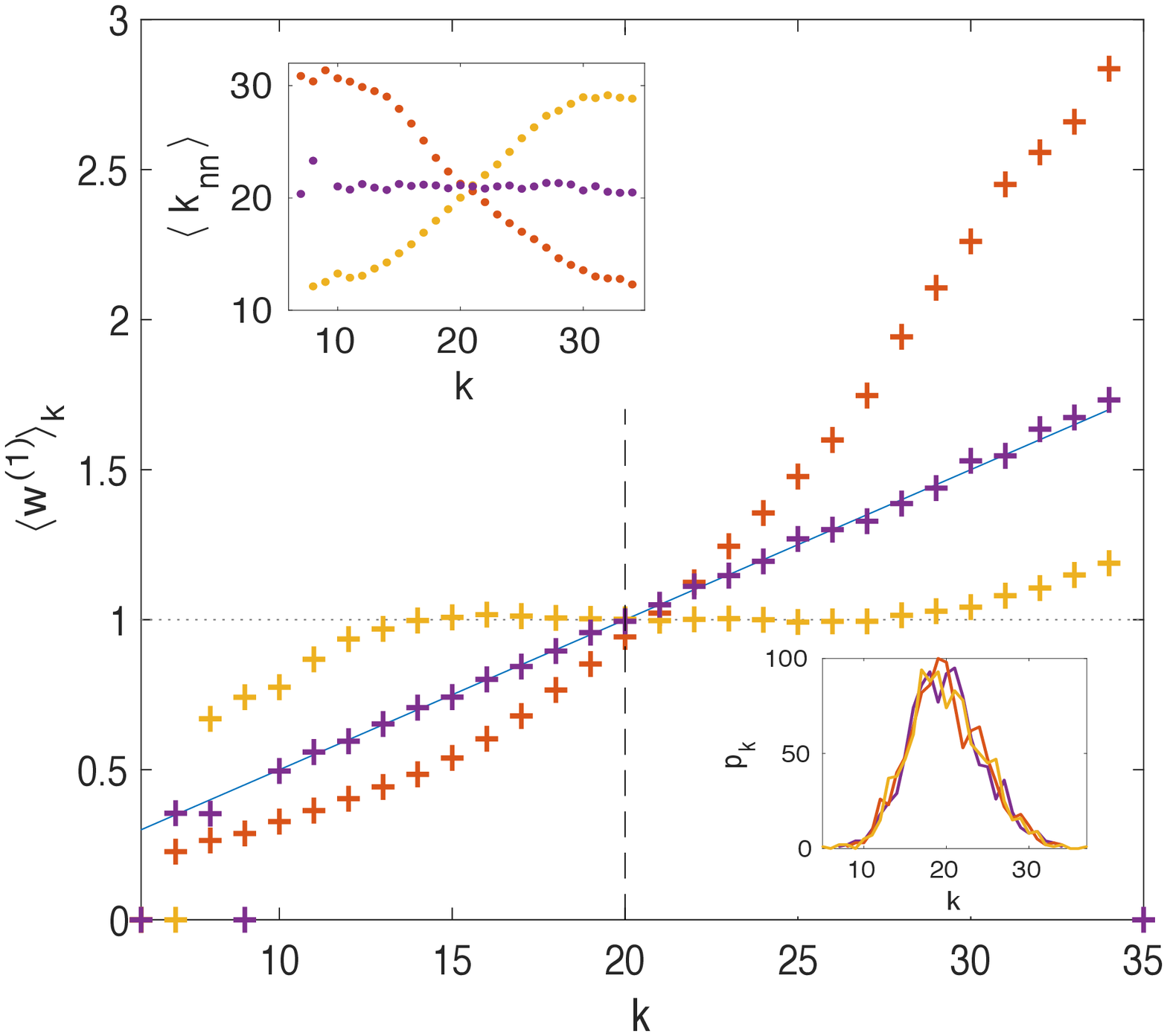}
	  \label{fig:degCorr_syntnot histogram}}
        \\
	\subfigure[]{\includegraphics[width=8.3cm]{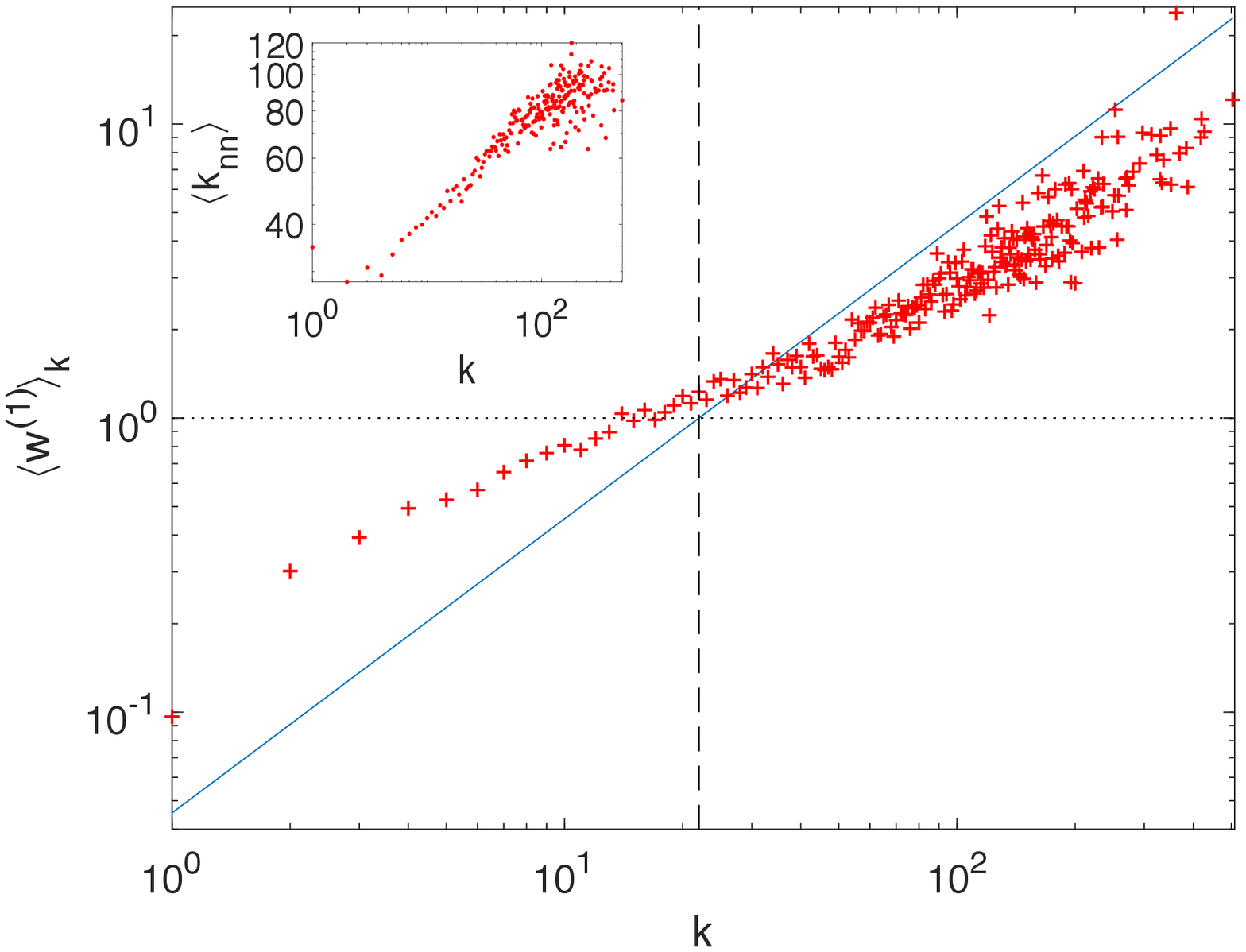}
		\label{fig:degCorr_astro}}
	\subfigure[]{\includegraphics[width=8.3cm]{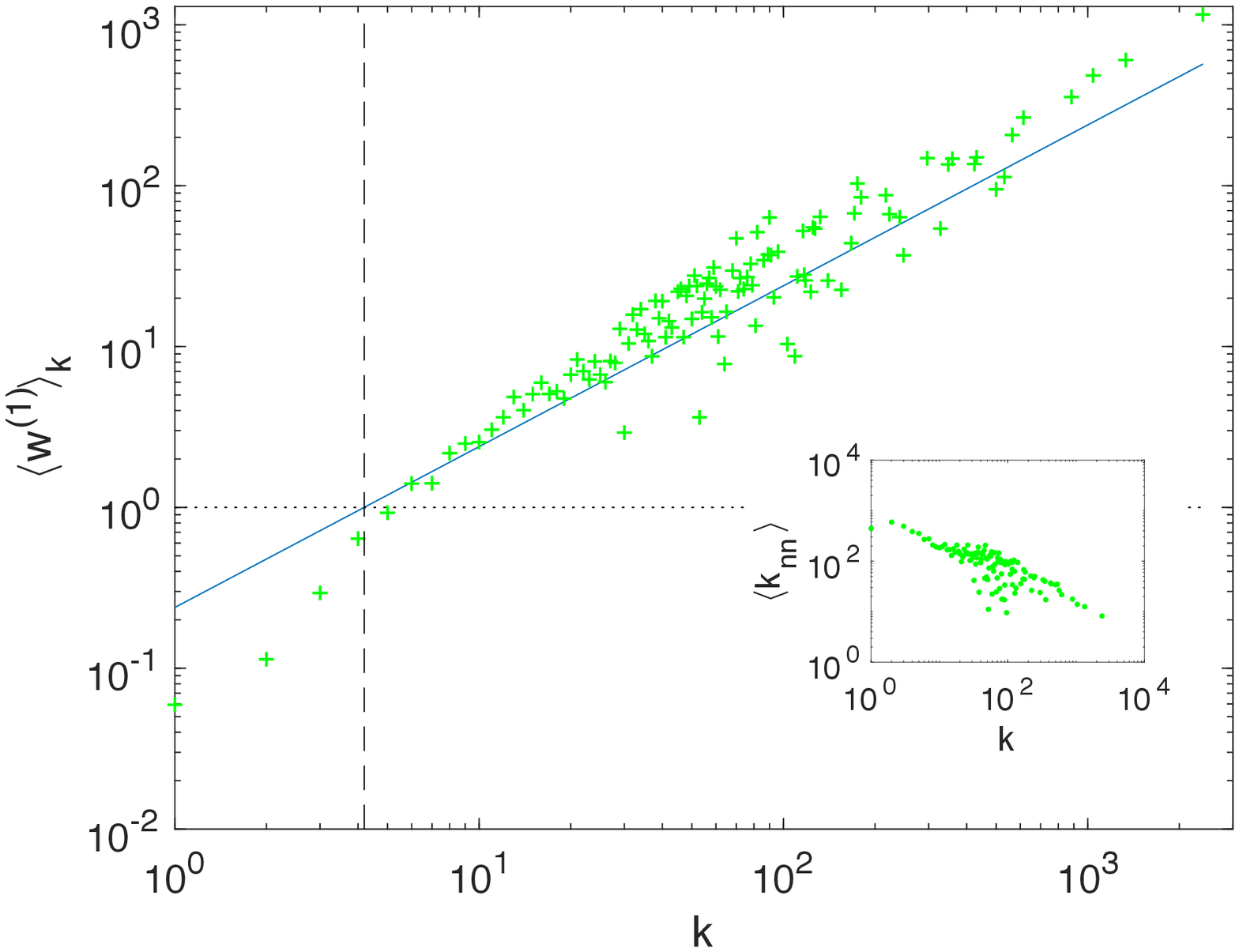}
		\label{fig:degCorr_Internet}}
\caption{The quantity $\langle w^{(1)} \rangle_k$ is reported as a
  function of $k$ for (a) three synthetic graphs with the same number
  of nodes and links, and respectively disassortative ($r=-0.94$, red
  pluses), assortative ($r=0.93$, yellow pluses) and uncorrelated ($r
  \sim 0$, purple pluses). The vertical dashed line identifies the
  mean degree $\langle k\rangle$. The mean degree  $\langle k_{nn}\rangle_k$
  of the nearest neighbours of nodes of degree $k$ is displayed in the
  upper-left inset, while the degree distribution $p_k$ is shown in
  the lower-right inset. Same quantities as in panel (a) for two real
  networks: (b) the network of collaborations in astrophysics and
  (c) Internet AS. Double-logarithmic scales have been used.}
	\label{fig:degCorr}
\end{figure}

In Fig. \ref{fig:degCorr}(a) we plot $\langle w^{(1)}\rangle_k$ as a
function of $k$ for three synthetic networks, respectively with
positive, negative and no degree correlations. The uncorrelated
network is an Erd\H{o}s-R\'enyi random graph with $N=1000$ nodes and
$K=10000$ edges, while the other two have been generated from the
uncorrelated one by using an algorithm that swaps edges according to
the degree of the corresponding
nodes~\cite{BonaventuraNicosiaLatora14,XulviBrunetSokolov05}, to
produce respectively a disassortative graph with correlation
coefficient $r=-0.94$ and an assortative graph with $r=0.93$ and
~\cite{Newman02}.  The algorithm preserves not only the average degree
$\avk$, but also the entire degree distribution, which is shown in
lower-right inset. Consequently, the results for the three networks,
shown respectively as purple, yellow and red pluses, can be directly
compared to the same analytical prediction $\langle w^{(1)}_{\rm nc}
\rangle_k$ (straight line), which is clearly well in agreement with
the randomized network.  In the disassortative graph we observe that
the quantity $\langle w^{(1)} \rangle_k$ is larger than $\langle
w_{\rm nc} \rangle_k$ for degree values $k> \avk$. This is
because the first neighbours of the hubs are typically poorly
connected, i.e. $\langle 1/k' \rangle_k > 1/\langle k\rangle$ when $k>
\avk$. Conversely, $\langle w^{(1)} \rangle_k$ is smaller than
$\langle w^{(1)}_{\rm nc} \rangle_k$ for poorly connected nodes,
i.e. for $k < \avk$.  In the assortative graph, as expected, $\langle
w \rangle_k \simeq 1$ for most of the degree classes. Deviations
from perfect assortativity only occur at the two extremes of the
degree distribution, i.e. for limit values of the degree: a sample
node with low (high) degree is in fact linked to nodes whose
degree is in average larger (lower) than its own.

\onecolumngrid

\twocolumngrid

In Fig. \ref{fig:degCorr}(b) and (c) we show the results obtained for
two real-world networks with known mixing patterns, namely the
collaboration networks of astrophysicists~\cite{Newman01} and the
Internet at the autonomous systems (AS) level
~\cite{PastorSatorrasVazquezVespignani01}. The first network has
$N=17903$, an average degree $\langle k \rangle$ equal to 22.2 and is
assortative with correlation coefficient $r=0.23$, while the second
one has $N=11174$, $\langle k \rangle=4.3$ and is disassortative with
$r=-0.19$. A logarithmic scale has been adopted in the two plots, as
both networks exhibit long-tailed degree distributions.  
The plots show larger fluctuations that those observed for the
artificially generated graphs. The general
behaviour is however preserved and allows to identify the two
different types of degree-degree correlations. In particular, the
inversion of the trend, which occurs around the mean degree, is clearly
preserved. For the assortative network of collaborations in astrophysics
$\langle w^{(1)} \rangle$ is larger than the value expected
for the uncorrelated case when 
$k<\langle k\rangle$, while it is smaller that this 
for almost all the larger values of $k$. The
opposite behaviour is displayed by the Internet network, which is instead 
disassortative. 
A possible way to detect the sign and, at the same time, to quantify the
entity of the correlations in a network from the study of the
quantity $\langle w^{(1)} \rangle_k$ is to extract 
the slope of the curve $\langle w^{(1)} \rangle_k$ as a function
of $k$ at point $k=\langle k\rangle$, and compare it to 
the slope of $\langle w^{(1)}_{\rm nc}\rangle_k$ vs $k$ for 
the corresponding randomized case.
For instance, we can evaluate the difference $\cal S$
between the two slopes multiplied by $\avk $:
\begin{eqnarray}
{\cal S} &=& \avk \frac{d}{dk}\Bigl(  \langle w^{(1)}_{\rm
	nc}\rangle_k - \langle w^{(1)}\rangle_k    \Bigr)\Big|_{k=\langle k\rangle} = \nonumber
\\
& = &  1 -\avk \frac{d}{dk}\biggl( k \sum_{k'} \frac{1}{k'} p_{k'|k} \biggr) \Bigg|_{k=\langle k\rangle} 
\end{eqnarray}
that we name {\em slope variation}. The multiplying mean degree has the role of rescaling $\cal S$, which becomes a quantity of order 1 (instead of $1/\avk$) and consequently a comparable measure for networks with different connectivity.
Such a quantity has been computed for the networks analysed in
Fig. \ref{fig:degCorr}. Results are reported in Table \ref{table:1}
and compared to the standard quantities usually adopted, namely 
the Pearson correlation coefficient $r$ and the
exponent $\nu$ governing the behaviour, $\langle k_{nn}\rangle_k \sim k^{\nu}$,
of the average degree of first neighbours of nodes of degree
$k$ as a function of $k$.
We notice that positive values of the slope variation $\cal S$
are associated to assortative networks, while negative slope differences
indicate disassortative ones, in agreement with the standard
indicators of degree-degree correlations.

Table \ref{table:1} also reports the values of $\cal S$ obtained in a sample of other artificial and real-world networks, and shows that the proposed indicator agrees not only for the sign but also for the order of magnitude with the standard measures, when evaluated for networks with strong degree-degree correlations, namely the network of collaboration in Astrophysics, Internet AS and Caida, as well as for artificial networks. The exceptional cases where the value of $\cal S$ results considerably different from $r$ and $\nu$ are those where the degree correlation does not prove to be clearly defined, corresponding to a significant error $\Delta \nu$ obtained from the fit of $k_{nn}(k)$.

In summary the value of $\cal S$ provides indication on the presence
of degree-degree correlations that are in all similar to $r$ or $\nu$.
However, as the $n$-th term of the expression of $\boldsymbol x^*$ in
Eq. \eqref{deltaxn} takes into account the degree correlations of a
node to those which are $n$ steps away, our indicator can be easily
generalized and employed to detect higher order degree correlations.
Let us consider for instance the second term $\delta x_i^{(2)}$ of the
Taylor expansion in Eq.\eqref{deltax2}. A second order analogue of
$w_i^{(1)}$ can be defined as $w_i^{(2)}\equiv\sum_{jl}
\frac{a_{ij}}{k_j}\frac{a_{jl}}{k_l}$ to measure the inverse degree of
the second neighbours of node $i$. Such a quantity represents a
measure of the connectivity of node $i$ compared to that of nodes
which are two steps away from it. The degree $k_j$ present at the
denominator mitigates the impact of the number of nodes adjacent to
$i$, so that the comparison only takes into account $k_i$ and the
degree of the second neighbours. Indeed, we have $w_i^{(2)}=1$ when all the
second neighbours $l$ of $i$ have degree $k_l=k_i$. As for the case of
$w_i^{(1)}$, we can consider the average value of $w_i^{(2)}$ over all
nodes $i$ of degree $k_i=k$. Writing this as a summation over degree
classes, we have:
\begin{equation}
\langle w^{(2)}\rangle_k = \frac{1}{N_k}\sum_{i=1}^N w_i^{(2)} \delta_{k_i,k}= k\sum_{k',k''}\frac{1}{k''} p_{k'|k} p_{k''|k'}
\label{w2}
\end{equation}
where $k''$ represents the degree of second neighbours. It is
important to notice that the above introduced quantity
does not measure genuine second order degree correlations in a network
but rather how the effect of first order degree correlations reflects on
nodes which are at distance of two steps. 
The generalization to higher orders follows naturally. 
Assessing the efficacy of this latter quantity 
as compared to other possible generalization of standard degree
correlation measures to higher order \cite{Allenperkins17} is left as a
challenge for future investigations.

\onecolumngrid

\vspace{5mm}
\begin{table*} [h]
	\begin{center}
		\begin{tabular}{|c|c|c|c|c|c|}
			\hline
			Networks &  $N$ & $\avk$  &       $r$ & $\nu\pm\Delta\nu$ & ${\cal S}$\\
			\specialrule{.15em}{.1em}{.1em} 
			\rowcolor{Gray}
			Synthetic uncorrelated & 1000 & 20 & -0.003 & -0.02 $\pm$ 0.01 & -0.01\\
			\hline
			\rowcolor{Gray}
			Synthetic assortative  & 1000 & 20 & 0.93 & 0.83 $\pm$ 0.08  & 1.02\\
			\hline
			 & 1000 & 20 & 0.71 & 0.61 $\pm$ 0.06 &  0.83\\
			 \hline
			 & 1000 & 20 & 0.50 & 0.34 $\pm$ 0.05 &  0.59 \\
			 \hline
			 & 1000 & 20 & 0.30 & 0.19 $\pm$ 0.03 &  0.36 \\
			\hline
			\rowcolor{Gray}
			Synthetic disassortative & 1000 & 20 & -0.94 & -0.89 $\pm$ 0.07  & -0.86\\
		        \hline
			 & 1000 & 20 & -0.71 & -0.66 $\pm$ 0.05 & -0.74\\
		        \hline
			 & 1000 & 20 & -0.50 & -0.35 $\pm$ 0.04 & -0.52\\
		        \hline
			 & 1000 & 20 & -0.30 & -0.23 $\pm$ 0.02 & -0.32\\
			\hline
			\hline
			Astrophysics collaboration~\cite{Newman01} & 17903 & 22.01 & 0.23 & 0.22 $\pm$ 0.02 & 0.41 \\
			\hline
			Facebook \cite{LeskovecMcauley12} & 4039 & 43.69 &0.11 & 0.054$\pm$ 0.051  & 0.40\\
			\hline
			Jazz collaboration\cite{jazz_net} & 198 & 27.70 & 0.03 & 0.11$\pm$ 0.04 & 0.46\\
			\hline
			Email URV \cite{Guimera_etal03}& 1134 & 9.61 & 0.078 & 0.05 $\pm$ 0.03&  0.03\\
			\hline
			C. elegans frontal\cite{KaiserHilgetag06} & 453 & 8.97 &0.035 & 0.062 $\pm$ 0.050 & 0.28\\
			\hline
			Internet AS \cite{PastorSatorrasVazquezVespignani01} & 11174 & 4.19 &-0.19 & -0.52 $\pm$ 0.04 & -0.33\\
			\hline
			Caida \cite{Leskovec_etal07} & 26475 & 4.03 & -0.19 & -0.52 $\pm$ 0.03 & -0.38\\
			\hline
			US politics books\cite{Krebs}  & 105 & 8.42 & -0.019 & -0.13 $\pm$ 0.07 &  -0.045\\
			\hline
			US power grid \cite{WattsStrogatz98} & 4941 & 2.67 & 0.003 & -0.035 $\pm$ 0.10 & -0.18\\
			\hline
		\end{tabular}
	\end{center}
\caption{Pearson correlation coefficient $r$, exponent $\nu$ and slope
  variation $\cal S$ for different synthetic and real-world networks with $N$ nodes
  and average degree $\langle k \rangle$. The highlighted rows correspond to the three artificial networks analysed in figure \ref{fig:degCorr_syntnot histogram}.}
	\label{table:1}
\end{table*}

\newpage
\twocolumngrid

\section{Conclusions}
\label{sec_concl}

Random walks have been extensively used to explore complex networks
with the aim of characterizing their structural features and unveil
their functional properties. In this
article we have introduced a class of random walkers that is subject
to node dependent reaction terms.  Our model of reactive random walks
is formulated in such a way that the relative contribution of the
interaction term at the nodes and of the relocation term can be tuned
at will, and this improves the sensitivity of the walkers to the
structure of the network.
In particular, the occupation probability of a given node is shaped by
the non trivial interplay between the connectivity patterns and the
local interaction functions. We have shown this by determining
analytically the asymptotic occupation probability via a perturbative
approach that takes a purely reactive dynamics as reference
point. Exploiting the dependence of the occupation probabilities on
the two tuning parameters of the model, namely the mobility parameter
$\mu$ and the bias parameter $\alpha$, and on the shape of the local
reaction functions, we have shown that reactive random walkers can
turn useful in many different ways.
We have first discussed how, by properly adjusting the reaction
contribution, one can emphasize nodes bridging otherwise disconnected
parts of the network, so that reactive random walkers can readily lead to
generalized definitions of node centrality measures. 
Furthermore, with the help of general arguments and of a 
series of worked examples we have shown that,   
by making the random walkers reactive and inspecting
their associated density distribution, one can easily
detect the symmetries of a network.
Finally, the specific form of the perturbative solution has inspired
the introduction of a novel indicator for the presence, sign and
entity of degree-degree correlations, which differently from other
standard measures is based on harmonic averages. We have illustrated
how reactive random walkers can distinguish assortative from
disassortative networks. The approach can in principle be
generalized to include next-to-leading correlations and this defines
an intriguing avenue for the investigation of higher-correlation in
complex networks which is left for future work.

In conclusion, we hope that this article has proven the versatility
and potential of reactive random walkers and that our work will trigger
further investigation of the model we have proposed and of its many
possible variations.

\section*{Acknowledgements}
V. L. acknowledges support from the EPSRC project EP/N013492/1.

\vfill	
\newpage
\section*{Appendix}

\subsection*{Random walks on networks} 
\label{appendix_A}

Random walks on networks are generally introduced as a discrete time
process governed by the equations
\begin{equation} 
x_i(n+1) = \sum_j \pi_{ij}x_j(n) 
\end{equation}   
where $x_i(n)$ denotes the probability that node $i$ is visited at
time step $n$. The stationary distribution
$\boldsymbol x^*=\lim_{n\rightarrow\infty}\boldsymbol x(n)$
satisfies the equation 
$\boldsymbol{x^*}= \Pi \boldsymbol{x^*}$ and, for undirected networks
$\pi_{ij} x_j^* = \pi_{ji}x_i^*$, meaning that the flow of probability
in each direction must equal each other at equilibrium
(\textit{detailed balance})~\cite{Sethna06}. This implies that, if
$\pi_{ij}=a_{ij}/k_j$, the stationary distribution is proportional to
the degree of nodes: $x_i^*=k_i/2K$.
\\
Switching from discrete to
continuous time when the spatial support is discrete, as in the case
of a network, is not trivial. The main point is to set the time scale
which is no longer simply defined by the discrete steps.  Two
different types of continuous-time random walks can be defined:
\textit{node-centric} and \textit{edge-centric}
~\cite{MasudaPorterLambiotte17}. In the node-centric version we
consider that a walker sitting on a node waits until the next move for
a time $\tau$, where $\tau$ is a random variable. If we assume that
there are independent, identical Poisson processes at each node of the
graph such that the walkers jump at a constant rate, the corresponding
continuous-time process is governed by:
\begin{equation*}
\dot x_i = \sum_j(\pi_{ij} - \delta_{ij})x_j\equiv \sum_j l^{\rm RW}_{ij}x_j.
\end{equation*}
where $L^{\rm RW} = \{ l^{\rm RW}_{ij} \}$, with $l^{\rm RW}_{ij} = \pi_{ij} - \delta_{ij}$ is the random walk Laplacian. 
The stationary state is then obtained by setting $\dot x_i$ equal to
zero, which gives $\sum_j l^{\rm RW}_{ij}x_j^*=0$, so yielding the
same stationary point of the discrete time version. This also
corresponds to the eigenvector of matrix $L^{\rm RW}$ associated to
eigenvalue $0$.
\\
In the other type of random walk, the edge-centric, which is
also generally called diffusion or fluid model
~\cite{AldousFill02,Samukhin_etal08,HoffmanPorterLambiotte12}, a step
occurs when the walker decides to move to another node by using one of
the outbound edges of its vertex, or in other words, when an edge is
activated. Clearly, the more connected is the starting node the larger
the set of options that can be alternatively selected to jump
away. The walker therefore leaves a node with large degree more
quickly than a node with small degree, and the transition rate for a walker starting
from node $i$ is equal to $k_i$. The occupation probability evolves in 
this case according to:
\begin{equation*}
\dot x_i = \sum_j(a_{ij} - k_j\delta_{ij})x_j\equiv \sum_j l^{\rm Diff}_{ij}x_j
\end{equation*}
which defines another Laplacian operator, $L^{\rm Diff} = \{ l^{\rm Diff}_{ij} \}$, with $l^{\rm Diff}_{ij} = a_{ij} - k_j\delta_{ij}$, associated to diffusion.
The stationary distribution $\boldsymbol x^*$ is in this case
homogeneous (as one would expect in a fluid model), being the
normalized eigenvector of $L^{\rm Diff}$ associated to 0 an
$N$-dimensional vector with all entries $1/N$.
\\
%
%

	\addcontentsline{toc}{chapter}{Bibliography}
	\bibliographystyle{ieeetr} 
	\bibliography{MyBib_gen}

\end{document}